\RequirePackage{fix-cm}
\documentclass{article}
\usepackage{graphicx}
\usepackage{authblk}
\usepackage{amssymb}
\usepackage{afterpage}
\usepackage[ruled,vlined]{algorithm2e}
\usepackage{amsmath}
\usepackage[nocompress]{cite}
\usepackage{comment}
\usepackage[bottom]{footmisc}
\usepackage{hyperref}
\usepackage{pgfplots}
\usepackage{subfig}
\usepackage{url}
\usepackage{tikz}
\usetikzlibrary{decorations.pathreplacing,patterns,backgrounds}
\usetikzlibrary{arrows,decorations.pathmorphing,backgrounds,positioning,fit,petri}
\usetikzlibrary{shapes}
\usepackage{wrapfig}

\newtheorem{definition}{Definition}
\newtheorem{example}{Example}

\AtBeginDocument{%
	\mathchardef\mathcomma\mathcode`\,
	\mathcode`\,="8000
}
{\catcode`,=\active
	\gdef,{\mathcomma\discretionary{}{}{}}
}
\mathchardef\breakingcomma\mathcode`\,
{\catcode`,=\active
	\gdef,{\breakingcomma\discretionary{}{}{}}
}

\newcommand{\abstrTyped}{\mathcal{A^{\typeAbstr}}}
\newcommand{\abstrTypedSub}[1]{\mathcal{A}_{#1}^{\typeAbstr}}
\newcommand{\abstrTypedUniv}{\mathcal{U_{A^{\typeAbstr}}}}

\newcommand{\abstractionFunctionLog}{\lambda^{\abstrTyped}_{L}}
\newcommand{\abstractionFunctionDataStrcuture}{\lambda^{\abstrTyped}_{\typedDataStructure}}
\newcommand{\activity}{A}

\newcommand{\caseIdent}{c}
\newcommand{\caseAct}{C \times A}
\newcommand{\caseUniv}{C}

\newcommand{\dataStructureOnStreamFunction}{\delta_{\typedDataStructure}}

\newcommand{\eventLog}{L}
\newcommand{\eventLogUniv}{\mathcal{U}_L}

\newcommand{\integersPos}{\mathbb{N}}
\newcommand{\integersPosZero}{\mathbb{N}_0}

\newcommand{\model}{\mathcal{M}}
\newcommand{\modelUniv}{\mathcal{U_{M}}}

\newcommand{\process}{\mathcal{P}}
\newcommand{\processDiscAlgo}{\gamma_L}
\newcommand{\processDiscAlgoAbstr}{\gamma^{\abstrTyped}}

\newcommand{\processUniv}{\mathcal{U_P}}

\newcommand{\sequence}{\sigma}
\newcommand{\stream}{\mathcal{S}}

\newcommand{\type}{\mathcal{T}}
\newcommand{\typeAbstr}{\mathbf{T}}
\newcommand{\typedDataStructure}{\mathcal{D^T}}
\newcommand{\typedDataStructureSub}[1]{\mathcal{D}_{#1}^{\mathcal{T}}}
\newcommand{\typedDataStructureUniverse}{\mathcal{U_{D^T}}}

\begin{document}
	
	\title{Event Stream-Based Process Discovery using Abstract Representations}
	
	\author{S.J. van Zelst\thanks{Corresponding Author: \texttt{s.j.v.zelst@tue.nl}}}
	\author{B.F. van Dongen}
	\author{W.M.P. van der Aalst}
	\affil{Department of Mathematics and Computer Science\\ 
		Eindhoven University of Technology\\ 
		P.O. Box 513, 5600 MB Eindhoven, The Netherlands\\ }

	\date{\today}

	\maketitle

\maketitle

\begin{abstract}
	The aim of process discovery, originating from the area of process mining, is to discover a process model based on business process execution data. 
	A majority of process discovery techniques relies on an event log as an input. 
	An event log is a static source of historical data capturing the execution of a business process.
	In this paper we focus on process discovery relying on online streams of business process execution events.
	Learning process models from event streams poses both challenges and opportunities, i.e. we need to handle unlimited amounts of data using finite memory and, preferably, constant time.
	We propose a generic architecture that allows for adopting several classes of existing process discovery techniques in context of event streams.
	Moreover, we provide several instantiations of the architecture, accompanied by implementations in the process mining tool-kit ProM\footnote{\url{http://promtools.org}}.
	Using these instantiations, we evaluate several dimensions of stream-based process discovery.
	The evaluation shows that the proposed architecture allows us to lift process discovery to the streaming domain.
%	\keywords{Process Mining \and Process Discovery \and Event Streams \and Abstract Representations}
\end{abstract}

\section{Introduction}
\label{sec:introduction}
\textit{Process mining}~\cite{DBLP:books/sp/Aalst16} aims at understanding and improving business processes.
The field consists of three main branches, i.e. \textit{process discovery}, \textit{conformance checking} and \textit{process enhancement}.
Process discovery aims at discovering a process model based on event data.
Conformance checking is concerned with assessing whether a process model and event data conform to each other in terms of possible behaviour.
Process enhancement is concerned with improvement of process models based on knowledge gained from event data, e.g. a process model is extended with performance diagnostics based on event data.

Several process discovery algorithms exist~\cite{aalst_2004_alpha,gunther_2007_fuzzy,weijters_2003_hm,leemans_2013_inductive_miner,aalst_2010_state_based_region,werf_2009_ilp}.
These algorithms all use an \textit{event log} as an input.
An event log is a static data source describing sequences of executed business process activities recorded over a historical time-span.
As the number of events recorded for operational processes is growing tremendously every year, so does the average event log size.
Conventional process discovery techniques are not able to cope with such large data sets, i.e. they fail when the data does not fit main memory.
Moreover, events are being generated at high rates, e.g. consider data originating from sensor networks, mobile devices and e-business applications.
Since existing process discovery techniques use static data, they are not able to capture the dynamics of such event streams in an adequate manner.

In this paper, we focus on process discovery using streams of business process events, i.e. \textit{event streams}, rather than event logs.
Applying process discovery on event streams allows us to gain insights in the underlying business process in a live fashion.
It furthermore allows us to deal with situations where: 
\textit{1.)} event logs are too large to fit main memory, 
\textit{2.)} there is no time to access event data continuously, i.e. real-time constraints and
\textit{3.)} recent behaviour is more important, i.e. concept drift.
A large class of existing process discovery algorithms transforms the event log into an \textit{abstract representation}, i.e. an abstraction of the event log, which is subsequently used to discover a process model.
To adopt these algorithms in a streaming context, it suffices to approximate the abstract representation based on the event stream.
Using abstract representations has several advantages:
\textit{1.)} \textit{Reusability}; We \textit{reuse} existing techniques by predominantly focusing on learning abstract representations from event streams.
\textit{2.)} \textit{Extensibility}; Once we design and implement a method for approximating a certain abstract representation, any (future) algorithm using the same abstract representation is automatically ported to event streams.
\textit{3.)} \textit{Anonymity}; In some cases, laws and regulations dictate that we are not allowed to store all event data. 
Some abstract representations ignore large parts of the data, effectively storing a summary of the actual event data, and therefore comply to \textit{anonymity} regulations.

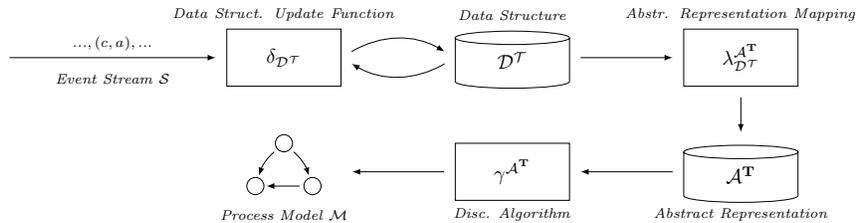
\begin{figure}[htb]
	\centering
	\resizebox{1\textwidth}{!}{
		\begin{tikzpicture}[
		>=latex,
		shorten >=2pt,
		shorten <=2pt,
		shape aspect=1,
		every place/.style= {minimum size=3mm},
		every transition/.style = {minimum size = 2mm}
		]
		
		\node (start) [] at (0,0) {};
		\node (stream) [] at (4,0) {};
		\draw[->] (start) -- (stream) node[midway,above]{\scriptsize $...,(c, a),...$};
		\node (stream_text) [] at (2,-0.375) {\scriptsize \textit{Event Stream} $\stream$};
		
		\draw (4,-0.5) rectangle (6,0.5);
		\node (ds_updater) [] at (5,0){$\dataStructureOnStreamFunction$};
		\node (ds_updater_text) [align=center,text width=5cm] at (5, 0.75) {\scriptsize \textit{Data Struct. Update Function}};
		\node (ds_update_from) [] at (6,0){};
		\node (ds_update_to) [] at (8,0){};

		\node (ds_text) [] at (9,0.75) {\scriptsize \textit{Data Structure}};
		\draw[->] (ds_update_from) to [bend left] (ds_update_to);
		\draw[->] (ds_update_to) to [bend left] (ds_update_from);
		
		\node (ds_label) [] at (9,-0.05) {$\typedDataStructure$};
		\node (ds) [cylinder, shape border rotate=90, draw, minimum height=1cm, minimum width=2cm] at (9,-0.125){};
		
		\node (ds_map_from) [] at(10,0){};
		\node (ds_map_to) [] at(12,0){};
		\draw[->] (ds_map_from) to (ds_map_to){};
		
		\draw (12,-0.5) rectangle (14,0.5);
		\node (mapper) [] at (13,0) {$\abstractionFunctionDataStrcuture$};
		\node (mapper_text) [align=center,text width=5cm] at (13, 0.75) {\scriptsize \textit{Abstr. Representation Mapping}};
		
		\node (mapper_from) [] at (13,-0.5){};
		\node (mapper_to) [] at (13, -1.5){};
		\draw [->] (mapper_from) to (mapper_to){};
		
		\node (abs) [cylinder, shape border rotate=90, draw, minimum height=1cm, minimum width=2cm] at (13,-2.125){};
		\node (abs_label) [] at (13, -2.05) {$\abstrTyped$};
		\node (abs_text) [] at (13, -2.75) {\scriptsize \textit{Abstract Representation}};
		
		\node (abs_from) [] at (12,-2){};
		\node (abs_to) [] at (10, -2) {};
		\draw[->] (abs_from) to (abs_to){};
		
		\draw (8,-2.5) rectangle (10,-1.5);
		\node (disc_label) [] at (9,-2) {$\processDiscAlgoAbstr$};
		\node (disc_text) [] at (9,-2.75) {\scriptsize \textit{Disc. Algorithm}};
		
		\node (disc_from) [] at (8,-2){};
		\node (disc_to) [] at (6,-2){};
		\draw[->] (disc_from) to (disc_to){};
		
		\node [place] (p1) [] at (5,-1.5) {};
		\node [place] (p2) [] at (5.5,-2.25) {}
		edge [pre, bend right = 15] node[auto] {} (p1);
		\node [place] (p3) [] at (4.5,-2.25) {}
		edge [pre, bend left = 15] node[auto] {} (p1)
		edge [pre] node[auto] {} (p2);
		\node[] (model text) at (5,-2.75) {\scriptsize \textit{Process Model} $\model$};
		\end{tikzpicture}
	}
	\caption{Schematic overview of the \textit{S-BAR architecture}.}
	\label{fig:architecture}
\end{figure}

We present the \textit{Stream-Based Abstract Representation (S-BAR)} architecture that describes this mechanism in a generic way (\autoref{fig:architecture}).
An event stream $\stream$ represents an (in)finite sequence of \textit{events}, emitted over time.
An \textit{event} is represented by a $(c,a)$-pair, stating that activity $a$ is executed in context of case $c$.
We maintain a data structure ($\typedDataStructure$) that represents the past behaviour emitted onto stream $\stream$.
Each time a new event arrives the data structure is kept up to date by updating its previous state based on the newly received event ($\dataStructureOnStreamFunction$).
From the data structure an \textit{algorithm-specific abstract representation} ($\abstrTyped$) is deduced ($\abstractionFunctionDataStrcuture$).
After learning the abstract representation, we reuse existing translations borrowed from conventional process discovery algorithms to return a process model $(\processDiscAlgoAbstr)$.

The S-BAR architecture is instantiated by designing a data structure, a data structure update mechanism and a data structure translation function.
The actual implementation of the data structure and related update functions influences the behaviour described by the discovered process model, e.g. using a time-decaying data structure versus a data structure that approximates the most frequent cases on the stream.
Several instantiations of the architecture have been implemented in the process mining toolkits ProM~\cite{dongen_2005_prom} and RapidProM~\cite{DBLP:journals/corr/AalstBZ2017,bolt_2015_swf}.
Using these implementations we conduct empirical experiments w.r.t. the behaviour of these algorithms in an event stream setting.
The experiments show that the algorithms are able to capture the behaviour reflected by the event stream.
Moreover, the experiments show that memory usage and processing times of the algorithms have non-increasing trends.

The remainder of this paper is organized as follows.
In \autoref{sec:preliminaries}, we present background information regarding business processes and process discovery.
In \autoref{sec:event_stream_discovery}, we present event streams and the notion of event stream based process discovery.
In \autoref{sec:abstr_repr}, we introduce the S-BAR architecture.
In \autoref{sec:instantiations}, we provide several instantiations of the architecture.
In \autoref{sec:evaluation}, we present an empirical evaluation of several instantiations of the architecture.
In \autoref{sec:related_work}, we present related work.
In \autoref{sec:discussion}, we discuss general challenges in event stream based process discovery.
\autoref{sec:conclusion} concludes the paper.

\section{Background}
\label{sec:preliminaries}
In this section we present general notation used throughout the paper and background concepts regarding business processes and process discovery.

$\integersPos$ denotes the set of positive integers, $\integersPosZero$ includes $0$.
A \textit{multiset} $B$ over set $X$ is a function $B : X \rightarrow \integersPosZero$.
We write a multiset as $[e_1^{k_1}, e_2^{k_2}, ..., e_n^{k_n}]$, where for $1 \leq i \leq n$ we have $e_i \in X$, $k_i \in \integersPos$ and $e_i^{k_i} \equiv B(e_i) = k_i$.
If for element $e$, $B(e) = 1$, we omit its superscript.
If for element $e$, $B(e) = 0$, we omit $e$ from the multiset notation.
An empty multiset is denoted as $[\ ]$. 
Element inclusion applies to multisets, i.e. if $e \in X$ and $B(e) > 0$ then $e \in B$.

A \textit{sequence} $\sigma$ of length $n$ relates positions to elements $e \in X$, i.e. $\sigma : \{1,2,...,n\} \rightarrow X$. 
An empty sequence is denoted as $\epsilon$. 
We write every non-empty sequence as $\langle e_1, e_2, ..., e_n \rangle$.
The set of all possible sequences over a set $X$ is denoted as $X^*$. 
We write \textit{concatenation} of sequences $\sigma_1$ and $\sigma_2$ as $\sigma_1 \cdot \sigma_2$.

Let $X,Y,Z$ and $Z'$ be sets and let $f: X \to Y$ and $g: Y \to Z$.
Function composition of $f$ and $g$ is defined as $g \circ f : X \to Z$, with $x \mapsto g(f(x))$ for $x \in X$.
Moreover, given $h: Z \to Z'$ we write $h \circ g \circ f$ for $h \circ (g \circ f)$, i.e. $h \circ g \circ f : X \to Z'$, with $x \mapsto h(g(f(x)))$ for $x \in X$.

\subsection{Business Processes, Models and Event Logs}
\begin{figure}[tb]
	\centering
	\includegraphics[height=0.65\textheight]{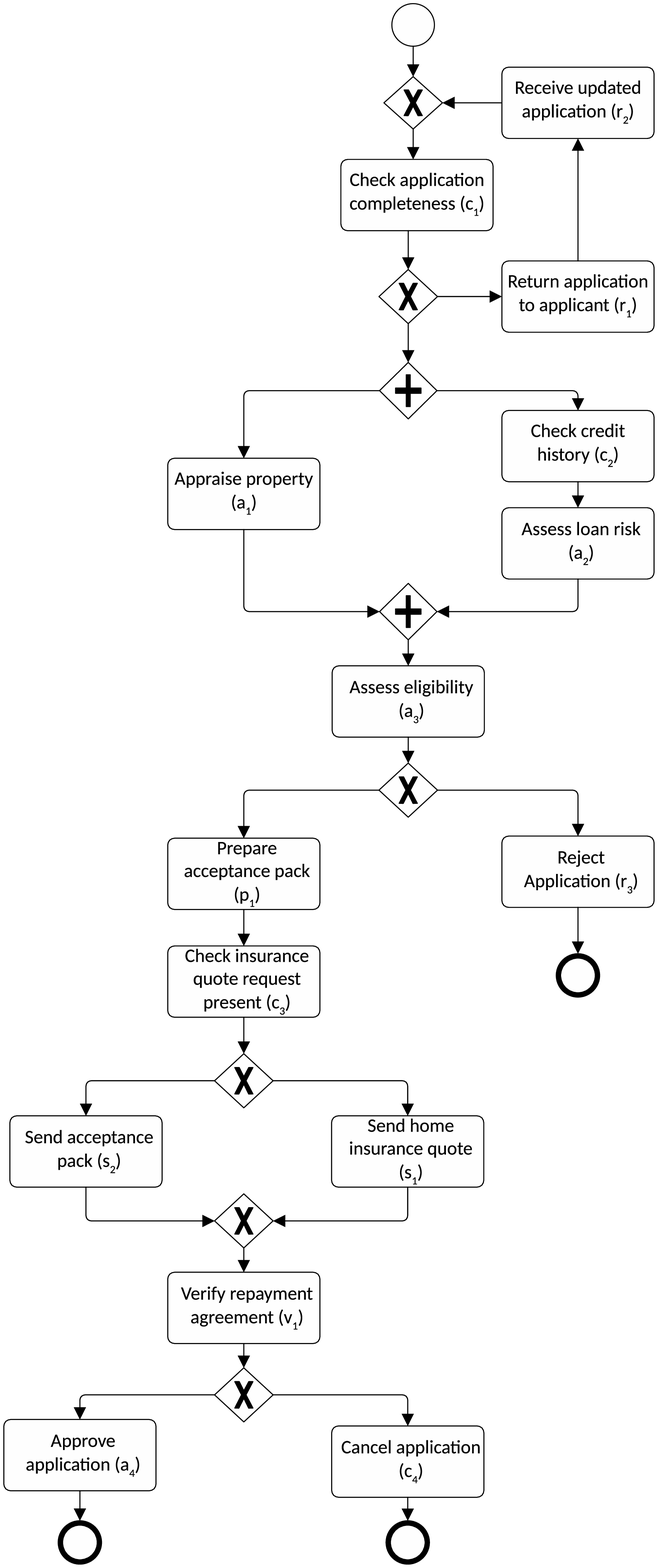}
	\caption{BPMN model of a loan application process (adopted from~\cite{dumas_2013_bpm_fundamentals}).}
	\label{fig:running_example_bpmn}
\end{figure}
\textit{Business processes} represent the execution of related business activities leading to a business goal.
Consider a bank offering loans to its customers.
A business goal of the bank is to accept, reject or cancel a loan application.
The bank's employees and its enterprise information system execute activities to achieve this goal, e.g. by checking a client's credit history and assessing the loan risk.

A business process $\process$ defines a set of sequences over a set of activities $\activity$, i.e. $\process \subseteq \activity^*$.
If $\sequence \in \process$ then the sequence of business activities $\sequence$ leads to a business goal and belongs to the \textit{behaviour} of $\process$.
In this paper, we assume the execution of activities to be atomic and abstract from data attributes such as resource, time-stamp etc.
Hence, we only consider the sequential ordering of activities (the \textit{control-flow perspective}).
$\processUniv$ denotes the universe of business processes.
A process model $\model$ \textit{represents} a business process and, like a process, defines a set of sequences over a set of activities $\activity$, i.e. $\model \subseteq \activity^*$.
$\modelUniv$ denotes the universe of process models.
In this paper, we consider process models that describe behaviour in a deterministic manner, e.g. Petri nets~\cite{murata_1989_petri_nets}, BPMN~\cite{omg_2011_bpmn} and workflow nets~\cite{aalst_1998_workflow_nets}.
Consider the BPMN model of a loan application handling process in \autoref{fig:running_example_bpmn}.
It describes that after an application is received, the first activity to be executed is \textit{``Check application completeness''}.
Depending upon the completeness of the application, the corresponding form is \textit{``Returned back to the applicant''}, or, the client's \textit{``credit history is checked''} and subsequently a \textit{``loan risk assessment''} is performed.
The two aforementioned activities can be executed concurrently with the \textit{``appraise property''} activity.
An \textit{``eligibility assessment''} of the loan is performed, eventually leading to a \textit{rejection}, \textit{cancellation} or \textit{approval} of the loan.

Today's information systems track the execution of business processes within a company.
Such systems store the execution of activities in context of a \textit{case}, i.e. an instance of the process.
The data stored by the information system is often in the form of an \textit{event log}.
Consider Table~\ref{tab:example_event_log} as an example.
\begin{table}[htb]
	\caption{Fragment of an event log.}
	\label{tab:example_event_log}
	\begin{center}
		\resizebox{0.85\textwidth}{!}{
			\begin{tabular}{|c|c|c|c|}
				\hline
				\textbf{Case} & \textbf{Activity} & \textbf{Resource} & \textbf{Time-stamp} \\
				\hline
				... & ... & ... & ...\\
				\hline
				\textit{3} & \textit{Approve application (a4)} & \textit{John} & \textit{2015-05-08 08:45} \\
				\hline
				\textit{4} & \textit{Check application completeness (c1)} & \textit{Lucy} & \textit{2015-05-08 09:13} \\
				\hline
				\textit{5} & \textit{Check application completeness (c1)} & \textit{John} & \textit{2015-05-08 09:14} \\
				\hline
				\textit{5} & \textit{Return application to applicant (r1)} & \textit{Pete} & \textit{2015-05-08 10:11} \\
				\hline
				\textit{5} & \textit{Receive updated application (r2)} & \textit{Pete} & \textit{2015-05-08 10:28} \\
				\hline
				\textit{6} & \textit{Check application completeness (c1)} & \textit{Lucy} & \textit{2015-05-08 10:33} \\
				\hline
				\textit{4} & \textit{Check credit history (c2)} & \textit{Rob} & \textit{2015-05-08 10:43} \\
				\hline
				\textit{5} & \textit{Appraise property (a1)} & \textit{Pete} & \textit{2015-05-08 11:00} \\
				\hline
				\textit{4} & \textit{Appraise property (a1)} & \textit{Rob} & \textit{2015-05-08 11:14} \\
				\hline
				\textit{4} & \textit{Assess loan risk (a2)} & \textit{Rob} & \textit{2015-05-08 11:35} \\
				\hline
				\textit{4} & \textit{Assess eligibility (a3)} & \textit{Lucy} & \textit{2015-05-08 11:55} \\
				\hline
				\textit{5} & \textit{Check credit history (c2)} & \textit{John} & \textit{2015-05-08 11:57} \\
				\hline
				\textit{4} & \textit{Prepare acceptance pack (p1)} & \textit{Lucy} & \textit{2015-05-08 12:25} \\
				\hline
				\textit{4} & \textit{Check insurance quote request present (c3)} & \textit{Lucy} & \textit{2015-05-08 12:23} \\
				\hline
				\textit{4} & \textit{Send acceptance pack (s2)} & \textit{Lucy} & \textit{2015-05-08 12:28} \\
				\hline
				\textit{5} & \textit{Assess loan risk (a2)} & \textit{John} & \textit{2015-05-08 12:37} \\
				\hline
				\textit{4} & \textit{Verify repayment agreement (v1)} & \textit{Lucy} & \textit{2015-05-09 13:05} \\
				\hline
				\textit{4} & \textit{Approve application (a4)} & \textit{John} & \textit{2015-05-09 14:15} \\
				\hline
				... & ... & ... & ...\\
				\hline
			\end{tabular}
		}
	\end{center}
\end{table}
The execution of an activity in context of a case, e.g. \textit{Approve application} executed for case \textit{3}, is referred to as an \textit{event}.
A sequence of events, e.g. the sequence of events related to case \textit{4}, $\langle$\textit{Check application form completeness, Check credit history, ..., Approve application}$\rangle$, is referred to as a \textit{trace} (written $\langle c_1, c_2, ...,a_4 \rangle$ when using abbreviated activity names).

An event log $\eventLog$ is a multiset of sequences over a set of activities $\activity$, i.e. $L : \activity^* \rightarrow \integersPosZero$, and describes the execution of some $\process \in \processUniv$.
$\eventLogUniv$ denotes the universe of event logs.
An event log is a \textit{sample} of the underlying process.
Therefore, there might exist process behaviour that is not present in the event log e.g., caused by parallelism.
In such case an event log is \textit{incomplete}.
There might also exist traces in the event log that are not part of the process, i.e. \textit{noisy} traces.
Noisy traces can be caused by faulty execution of the process, incomplete specifications or technical issues such as incorrect logging, system errors and mixed time granularity.

\subsection{Process Discovery}
\label{subsec:prelim_process_discovery}
The goal of Process discovery is to discover a process model based on an event log.
Several process discovery algorithms exist~\cite{aalst_2004_alpha,gunther_2007_fuzzy,weijters_2003_hm,leemans_2013_inductive_miner,aalst_2010_state_based_region,werf_2009_ilp}.
These algorithms differ in terms of their underlying computational schemes and data structures as well as their resulting process modeling formalism.
We refer to~\cite{DBLP:books/sp/Aalst16,dongen_2009_overview, weerdt_2012_discovery_algorithms} for a detailed overview of process discovery algorithms.

A process discovery algorithm $\processDiscAlgo$ discovers a process model based on an event log, i.e. $\processDiscAlgo : \eventLogUniv \to \modelUniv$.
The challenge is to design $\processDiscAlgo$ in such way that $\processDiscAlgo(\eventLog)$ is an \textit{appropriate representation} of the underlying process $\process$.
Appropriateness of $\gamma(L)$ depends on the aim of the process discovery analysis, e.g. ensuring that all behaviour in the event log is present in the model versus ensuring that the most frequent behaviour is present.
Given the different aims of process discovery analyses, several quality measures are defined in order to judge their resulting model's appropriateness.
Ideally, $\process$ is used as a basis to compute these metrics, however, as $\eventLog$ is the only tangible sample of $\process$, we typically compute the quality of $\processDiscAlgo(\eventLog)$ using $\eventLog$.
The four essential process mining quality dimensions are \textit{replay fitness}, \textit{precision}, \textit{simplicity} and \textit{generalization}~\cite{DBLP:books/sp/Aalst16,buijs_2014_qd}.
Replay fitness describes what fraction of the behaviour present in $\eventLog$ is also described by $\processDiscAlgo(\eventLog)$.
Precision describes what fraction of the behaviour described by $\processDiscAlgo(\eventLog)$ is also present in $\eventLog$.
Simplicity describes the (perceived) complexity of the process model.
Since it is unlikely that the event log contains all behaviour (incompleteness), generalization describes how well the process model generalizes for behaviour not present in $\eventLog$.
Due to noise, an algorithm guaranteeing perfect replay fitness, i.e. all behaviour in the event log is present in the discovered model, captures behaviour that is not part of the process.
In practice this leads to very complex models that are impossible to be interpreted by a human analyst.
Hence, a process discovery algorithm needs to strike \textit{an adequate balance} between the four essential quality dimensions.

\section{Event Stream Based Process Discovery}
\label{sec:event_stream_discovery}
Existing process discovery techniques discover process models in an a-posteriori fashion, i.e. they provide a historical view of the data.
However, most information systems allow us to capture the execution of activities at the moment they occur.
Discovering and analysing process models from such continuous streams of events allows us to get a real-time view of the process under study.
Such view paves the way for new types of process mining analysis, i.e. we are able to answer more advanced questions such as ``What is the current status of the process?'' and ``What running cases are likely to cause problems?''.
It also allows us to inspect and visualize recent behaviour and evolution of behaviour in the process, i.e. concept drift.

There are several other advantages of studying streams of events rather than event logs.
Trends such as \textit{Big Data} and \textit{Data Science} signify the spectacular growth and omnipresence of data.
Typically, real event logs do not fit main memory.
Since we assume event streams to be potentially infinite, analysing them enables us to handle event data of arbitrary size.
In other cases we do not have the time or are not allowed to access event data continuously and, hence, need to analyse events at the moment they occur.

In this section we formalize \textit{event streams} and \textit{event stream based process discovery}.
Additionally we quantify high-level requirements for the design of event stream based process discovery algorithms.

\subsection{Event Streams}
An event stream is a continuous stream of events executed in context of an underlying business process.
We represent an event stream as a sequence of  pairs consisting of a \textit{case-identifier} and an \textit{activity}.
Hence, for each event we know what activity was performed in context of what process instance.
When comparing event streams to event logs, we identify two main differences:
\textit{1.)} an event stream is potentially \textit{infinite} and \textit{2.)} behaviour seen for a case is \textit{incomplete}, i.e. in the future new events may be executed in context of a case.

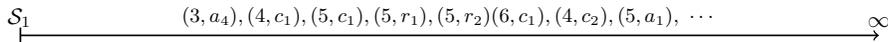
\begin{figure}[tb]
	\resizebox{1\textwidth}{!}{
		\begin{tikzpicture}
		\draw[|->, line width=0.75pt] (-6.75,0) node[anchor=south]{$\stream_1$} -- (6.75,0) node[anchor=south]{$\infty$};
		\draw (0,0) node[anchor=south]{\small $(3,a_4),(4,c_1),(5,c_1),(5,r_1),(5,r_2)(6,c_1),(4,c_2),(5,a_1),\ \cdots$};
		\end{tikzpicture}
	}
	\caption{Example event stream $\stream_1$.}
	\label{fig:example_stream}
\end{figure}

\begin{definition}[Event stream]
	\label{def:event_stream}
	Let $\activity$ be a set of activities and let $\caseUniv$ denote the set of all possible case identifiers.
	An event stream $\stream$ is a sequence over $\caseAct$, i.e. $\stream \in (\caseAct)^*$.
\end{definition}

A pair $(\caseIdent,a) \in \caseAct$ represents an event, i.e. activity $a$ was executed in context of case $\caseIdent$.
$\stream(1)$ denotes the first event that we receive, whereas $\stream\textit{(i)}$ denotes the $i^{th}$ event.
Consider stream $\stream_1$ in \autoref{fig:example_stream} as an example, where, event $(3,a_4)$ is emitted first ($\stream_1(1) = (3,a_4)$), event $(4,c_1)$ is emitted second and event $(5,a_1)$ is the eight and last event emitted onto the stream \textit{up until now}.
We receive multiple events related to the same case at different points in time, e.g. the second and seventh event on $\stream_1$ are related to case $4$.
Hence, handling such type of data needs new types of data structures and event processing techniques compared to conventional process discovery.

\subsection{Process Discovery}
\label{subsec:stream_process_discovery}
The goal of event stream based process discovery is to discover a process model using an event stream as an input.
A first step is to approximate, based on $\stream$, the presence of some $\sigma \in \process$ and possibly $\sigma$'s frequency w.r.t. $\stream$.
Given such approximation the next step is to deploy a process discovery algorithm onto the approximation in order to obtain a process model.

A naive approach is to construct an event log based on the event stream by using a data structure that stores case-sequence pairs $(\caseIdent, \sigma) \in C \times \activity^*$.
For every event $(\caseIdent, a)$ we receive, we check whether the data structure contains entry $(\caseIdent,\sigma')$.
If so, we update this entry to $(\caseIdent, \sigma' \cdot \langle a \rangle)$.
If not, we insert new entry $(\caseIdent, \langle a \rangle)$.
Whenever we want to discover a new process model based on the current state of the event stream, we transform the data structure into an event log and provide it to any conventional process discovery algorithm.
Observe that, since the stream is potentially infinite, this procedure needs \textit{infinite memory}.
Moreover, the approach includes \textit{redundancy}, i.e. several (partial) traces that where already analysed in a previous call to a discovery algorithm, and are still in memory at the next call, are analysed twice.
Hence, we want the data structure to either represent, or be easily translatable to, some minimal form of data needed in order to discover a process model.

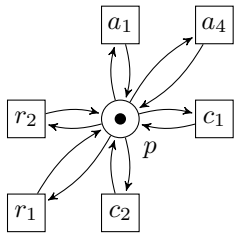
\begin{wrapfigure}{L}{0.425\textwidth}
	\centering
	\begin{tikzpicture}	
	[   node distance=1.25cm,
	on grid,>=stealth',
	bend angle=15,
	auto,
	every place/.style= {minimum size=5mm},
	every transition/.style = {minimum size = 5mm}
	]
	
	\node [place, tokens = 1] (p) [label=below right:$p$]{};
	
	\node [transition] (a) [label=center:$a_1$, above = of p] {}
	edge [pre, bend right = 15] node[auto] {} (p)
	edge [post, bend left = 15] node[auto] {} (p);
	
	\node [transition] (a4) [label=center:$a_4$, right = of a] {}
	edge [pre, bend right = 15] node[auto] {} (p)
	edge [post, bend left = 15] node[auto] {} (p);

	\node [transition] (c1) [label = center:$c_1$, right = of p] {}
	edge [pre, bend right = 15] node[auto] {} (p)
	edge [post, bend left = 15] node[auto] {} (p);
	
	\node [transition] (c2) [label=center:$c_2$,  below = of p] {}
	edge [pre, bend right = 15] node[auto] {} (p)
	edge [post, bend left = 15] node[auto] {} (p);
	
	\node [transition] (r1) [label = center:$r_1$, left = of c2] {}
	edge [pre, bend right = 15] node[auto] {} (p)
	edge [post, bend left = 15] node[auto] {} (p);
	
	\node [transition] (d) [label = center:$r_2$, left = of p] {}
	edge [pre, bend right = 15] node[auto] {} (p)
	edge [post, bend left = 15] node[auto] {} (p);
	\end{tikzpicture}
	\caption{Example ``flower'' model.}
	\label{fig:flower}
	\vspace{-10pt}
\end{wrapfigure}

An example of an algorithm using a minimal data representation is the \textit{flower-miner}.
The flower-miner produces a process model that allows for every possible sequence over the observed activities.
Reconsider example stream $\stream_1$ (\autoref{fig:example_stream}) which consists of activities labelled $a_1$, $a_4$, $c_1$, $c_2$, $r_1$ and $r_2$.
In \autoref{fig:flower} we depict a flower model, in terms of a Petri net~\cite{murata_1989_petri_nets}, that allows for all activities on $\stream_1$.

To ensure that the flower miner uses finite memory, we just need to deploy any finite memory based data structure that keeps track of the activities seen on the stream.
A wide variety of such data structures exits, e.g. count-based frequent item data structures~\cite{cormode_2009_freq_item_methods}, reservoirs~\cite{vitter:1985:_reservoir, aggarwal_2006_biased_rs} and time-decay based models~\cite{cormode_2009_forward_decay}.
Whenever we receive a new event $(\caseIdent, a)$ we just add $a$ to the data structure.
Translating the data structure to a process model is trivial, i.e. every activity present in the data-structure is adopted in the flower model.

The flower miner works, yet it has deficiencies from a process discovery perspective.
It generalizes the behaviour represented by the event stream as much as possible.
The resulting process model very likely allows for \textit{much more behaviour} than actually present in the underlying process.
Hence, we need techniques that are \textit{more precise}.

The event log based approach and the flower miner represent two extremes.
Storing the event stream as an event log requires us to reuse a large part of the data several times.
The flower miner on the other hand neglects a large quantity of information carried by the event stream and greatly over-generalizes the stream's behaviour.
We therefore need a scheme that is in the middle of both extremes, i.e. it does not store the complete event log, yet it stores enough data to provide meaningful output.

\section{The S-BAR Architecture}
\label{sec:abstr_repr}
When analysing conventional process discovery algorithms, we observe that a majority shares a common underlying algorithmic mechanism.
The event log is transformed into an \textit{abstract representation}, which is subsequently used to construct a resulting process model.
Moreover, several algorithms use the same abstract representation.
In \autoref{example:alpha} we illustrate the \textit{directly follows abstraction}, used by the \textit{$\alpha$-Miner}~\cite{aalst_2004_alpha}.

\begin{example}[The directly follows abstraction \& the $\alpha$-Miner]
	\label{example:alpha}
	Consider event log $L = [\langle a,b,c,d \rangle, \langle a,c,b,d \rangle]$. 
	The $\alpha$-Miner computes a \textit{directly follows abstraction} based on the event log.
	Activity $a$ is directly followed by $b$, written as $a > b$, if there exists some sequence $\sigma \in L$ of the form $\sigma = \sigma' \cdot \langle a,b \rangle \cdot \sigma''$.
	In case of event log $L$ we deduce $a > b$, $a > c$, $b > c$, $b > d$, $c > b$, $c > d$.
	Using these relations as a basis, the $\alpha$-Miner constructs a Petri net.
\end{example}

As \autoref{example:alpha} shows, the event log is translated into a \textit{directly follows abstraction}, which is subsequently used to construct a process model.
Other discovery algorithms like the Inductive Miner~\cite{leemans_2013_inductive_miner} and the ILP Miner~\cite{werf_2009_ilp} use \textit{the same mechanism} to discover a process model.
To adopt these algorithms to an event stream context, it suffices to determine whether we are able to learn the corresponding abstract representation from the event stream and, if possible, design a data structure that supports this.

In the remainder of this section we formalize the notion of abstract representations.
Subsequently we introduce the \textit{Stream-Based Abstract Representation} (S-BAR) architecture that captures the notion of event stream based abstract representation computation in a generic manner.

\subsection{Abstract Representations in Conventional Process Discovery}
We refine conventional process discovery by splitting $\processDiscAlgo$ into two steps.
In the first step, the event log is translated into the abstraction used by the discovery algorithm.
In the second step, the abstraction is translated into a process model.
In the remainder we let $\typeAbstr$ denote an abstract representation type.
$\abstrTyped$ denotes an abstract representation of type $\typeAbstr$ and $\abstrTypedUniv$ denotes the universe of abstract representations of type $\typeAbstr$.
\begin{definition}[Abstraction Function - Event Log]
	\label{def:abstraction_funtion_event_log}
	Let $\typeAbstr$ denote an abstract representation type.
	An abstraction function $\abstractionFunctionLog$ is a function mapping an event log to an abstract representation of type $\typeAbstr$.
	\begin{equation}
	\abstractionFunctionLog \colon \eventLogUniv \to \abstrTypedUniv
	\end{equation}
\end{definition}

Using \autoref{def:abstraction_funtion_event_log}, we define process discovery in terms of abstract representations.
\begin{definition}[Process Discovery Algorithm - Abstract Representation]
	\label{def:process_discovery_algorithm_abstract_representation}
	Let $\typeAbstr$ denote an abstract representation type.
	An abstract representation based process discovery algorithm $\processDiscAlgoAbstr$ maps an abstract representation of type $\typeAbstr$ to a process model.
	\begin{equation}
	\processDiscAlgoAbstr \colon \abstrTypedUniv \to \modelUniv
	\end{equation}
\end{definition}

\begin{figure}[tb]
	\centering
	\resizebox{0.85\textwidth}{!}{
		\begin{tikzpicture}[
		>=latex,
		shorten >=2pt,
		shorten <=2pt,
		shape aspect=1,
		every place/.style= {minimum size=6mm},
		every transition/.style = {minimum size = 6mm}
		]
		\node (L) [cylinder, shape border rotate=90, draw, minimum height=1cm, minimum width=2cm] at (0,0) {};
		\node[] (log_text) at (0,-1) {\scriptsize \textit{Event Log}};
		\node[] (log_label) at (0,-0.05) {$\eventLog \in \eventLogUniv$};
		
		\node[] (abs) [] at (2.125, 0.375) {$\abstractionFunctionLog$};
		\draw[->, dashed] (1.25,0) to (3,0);
		
		\node (C1) [fill,shape=circle,inner sep=0pt,minimum size=2.5pt, label=below left:\scriptsize$c_1$] at (4,0) {};
		\node (R1) [fill,shape=circle,inner sep=0pt,minimum size=2.5pt, label=right:\scriptsize$r_1\ \cdots$] at (4.5,0.5) {};
		\node (R2) [fill,shape=circle,inner sep=0pt,minimum size=2.5pt, label=left:\scriptsize$r_2$] at (4,0.5) {};
		\node (A1) [fill,shape=circle,inner sep=0pt,minimum size=2.5pt, label=right:\scriptsize$a_1\ \cdots$] at (4.5,0) {};
		\node (C2) [fill,shape=circle,inner sep=0pt,minimum size=2.5pt, label=right:\scriptsize$c_2\ \cdots$] at (4.5,-0.5) {};
		\node[] (log_label) at (5,-1) {\scriptsize\textit{Directly Follows Abstraction}};
		
		\draw[->] (C1) to (R1);
		\draw[->] (C1) to (A1);
		\draw[->] (C1) to (C2);
		\draw[->] (R1) to (R2);
		\draw[->] (R2) to (C1);
		
		\node[] (abs) [] at (7.625, 0.375) {$\processDiscAlgoAbstr$};
		\draw[->, dashed] (6.75,0) to (8.5,0);
		
		\node [place, tokens = 1] (p1) [] at (9,0) {};
		\node [transition] (t) [right = of p1] {}
		edge [pre, bend left = 15] node[auto] {} (p1)
		edge [post, bend right = 15] node[auto] {} (p1);
		
		\node[] (log_label) at (10,-1) {\scriptsize \textit{Process Model}};
		\end{tikzpicture}
	}
	\caption{The $\alpha$-Miner in terms of its abstract representation.}
	\label{fig:two_phase_discovery_alpha}
\end{figure}
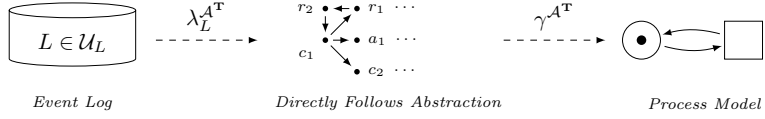

Every discovery algorithm that uses an abstract representation internally can be expressed as a composition of $\abstractionFunctionLog$ and $\processDiscAlgoAbstr$.
Thus, given event log $\eventLog \in \eventLogUniv$ and abstract representation type $\typeAbstr$,  we obtain $\processDiscAlgo = (\processDiscAlgoAbstr \circ \abstractionFunctionLog)(\eventLog)$.
For example, consider \autoref{fig:two_phase_discovery_alpha} depicting the $\alpha$-Miner in terms of $\processDiscAlgoAbstr$ and $\abstractionFunctionLog$.

\subsection{Abstract Representations in Event Stream Based Process Discovery}
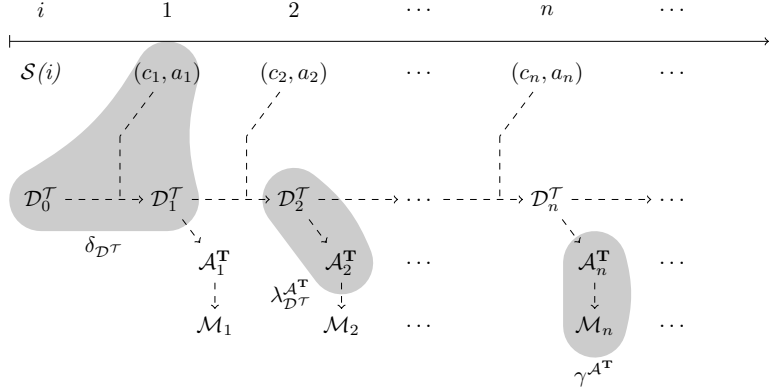
\begin{figure}[tb]
	\centering
	\resizebox{0.85\textwidth}{!}{
		\begin{tikzpicture}
		%COL: 1
		\node[] (t) at (0,0) {$i$};
		\node[] (stream) at (0,-1) {$\stream\textit{(i)}$};
		
		%Timeline
		\draw[|->] (-0.5,-0.5) to (11.5,-0.5);
		
		\node[] (ds10) at (0,-3) {$\typedDataStructureSub{0}$};
		
		%data structure update function
		\node[] at (1,-3.75) {$\dataStructureOnStreamFunction$};
		
		%COL: 2
		\node[] at (2,0) {$1$};
		\node[] (e1) at (2,-1) {$(\caseIdent_1, a_1)$};
		
		\node[] (ds11) at (2,-3) {$\typedDataStructureSub{1}$};
		\draw[->, dashed] (ds10) to (ds11);
		
		\draw[-, dashed] (e1) to (1.25,-2);
		\draw[-, dashed] (1.25,-2) to (1.25,-3) ;
		
		%COL: 2.5
		\node[] (a11) at (2.75,-4) {$\abstrTypedSub{1}$};
		\draw[->, dashed] (ds11) to (a11);
		\node[] (m11) at (2.75,-5) {$\model_1$};
		\draw[->, dashed] (a11) to (m11);

		%COL: 3
		\node[] at (4,0) {$2$};
		\node[] (e2) at (4,-1) {$(\caseIdent_2, a_2)$};
		
		\node[] (ds12) at (4,-3) {$\typedDataStructureSub{2}$};
		\draw[->, dashed] (ds11) to (ds12);
		
		\draw[-, dashed] (e2) to (3.25,-2);
		\draw[-, dashed] (3.25,-2) to (3.25,-3);
		
		\node[] (dsAbs) at (3.95,-4.5) {$\abstractionFunctionDataStrcuture$};
		
		%COL: 3.5
		\node[] (a12) at (4.75,-4) {$\abstrTypedSub{2}$};
		\draw[->, dashed] (ds12) to (a12);
		\node[] (m12) at (4.75,-5) {$\model_2$};
		\draw[->, dashed] (a12) to (m12);
		
		%COL: 4
		\node[] at (6,0) {$\dots$};
		\node[] at (6,-1) {$\dots$};
		
		\node[] (ds1dots) at  (6,-3) {$\dots$};
		\draw[->, dashed] (ds12) to (ds1dots);
		
		\node[] at  (6,-4) {$\dots$};
		\node[] at  (6,-5) {$\dots$};
		
		%COL: 5
		\node[] at (8,0) {$n$};
		\node[] (em) at (8,-1) {$(\caseIdent_n, a_n)$};
		
		\node[] (ds1m) at (8,-3) {$\typedDataStructureSub{n}$};
		\draw[->, dashed] (ds1dots) to (ds1m);
		
		\draw[-, dashed] (em) to (7.25,-2);
		\draw[-, dashed] (7.25,-2) to (7.25,-3);
		
		%COL: 5.5
		\node[] (a1m) at (8.75,-4) {$\abstrTypedSub{n}$};
		\draw[->, dashed] (ds1m) to (a1m);
		\node[] (m1m) at (8.75,-5) {$\model_n$};
		\draw[->, dashed] (a1m) to (m1m);
		
		\node[] (absDisc) at (8.75,-5.75) {$\processDiscAlgoAbstr$};

		%COL: 6
		\node[] at (10,0) {$\dots$};
		\node[] at (10,-1) {$\dots$};
		
		\node[] (ds1dots2) at  (10,-3) {$\dots$};
		\draw[->, dashed] (ds1m) to (ds1dots2);
		
		\node[] at  (10,-4) {$\dots$};
		\node[] at  (10,-5) {$\dots$};
		
		%from http://tex.stackexchange.com/questions/27171/padded-boundary-of-convex-hull?lq=1
		\begin{pgfonlayer}{background}
		\foreach \nodename in {e1,ds10,ds11} {
			\coordinate (\nodename') at (\nodename);
		}
		\path[fill=black!20,draw=black!20,line width=1cm, line cap=round, line join=round] 
		(e1') to[bend left=15] (ds10') 
		to(ds11')[bend left=20]
		to[bend left=15] (e1') -- cycle;
		\end{pgfonlayer}
		
		\begin{pgfonlayer}{background}
		\foreach \nodename in {ds12,a12} {
			\coordinate (\nodename') at (\nodename);
		}
		\path[fill=black!20,draw=black!20,line width=1cm, line cap=round, line join=round] 
		(ds12') to[bend left=15] (a12') 
		to(ds12')[bend left=20] -- cycle;
		\end{pgfonlayer}
		
		\begin{pgfonlayer}{background}
		\foreach \nodename in {a1m,m1m} {
			\coordinate (\nodename') at (\nodename);
		}
		\path[fill=black!20,draw=black!20,line width=1cm, line cap=round, line join=round] 
		(a1m') to[bend left=15] (m1m') 
		to(a1m')[bend left=20] -- cycle;
		\end{pgfonlayer}
		\end{tikzpicture}
	}
	\caption{Detailed overview of the S-BAR architecture.}
	\label{fig:architecture2}
\end{figure}

In this section we present the S-BAR architecture which captures the use of abstract representations in an event stream context in a generic manner.
In \autoref{fig:architecture2} the S-BAR architecture is depicted schematically.
The S-BAR architecture conceptually splits event-stream-based process discovery into three components, highlighted in gray in \autoref{fig:architecture2}.
We explain the purpose of each component, i.e.  $\dataStructureOnStreamFunction$, $\abstractionFunctionDataStrcuture$ and $\processDiscAlgoAbstr$, by means of an example.

Consider maintaining the directly follows abstraction, introduced in \autoref{example:alpha}, on a stream.
To do this, we need a data structure that tracks the most recent activity for each case.
Given such data structure, if we receive new event $(c,a)$, we check whether we already received an activity $a'$ for case $c$ or whether $a$ is the first activity received for case $c$.
If we already received activity $a'$ for case $c$, we deduce $a' > a$.
Subsequently we update our data structure such that it now assigns $a$ to be the last activity received for case $c$.

The first component, i.e. $\dataStructureOnStreamFunction$, maintains and updates a (collection of) data structure(s) that together form a \textit{sufficient representation} of the behaviour entailed by the event stream.
In context of our example, the first component is mainly concerned with keeping track of pairs of activities that are in a $a' > a$ relation.
The second component, i.e. $\abstractionFunctionDataStrcuture$, translates the data structure to an abstract representation.
In context of our example, this consists of translating the pairs of activities that are in a $a' > a$ relation into the directly follows abstraction.
The third component, i.e. $\processDiscAlgoAbstr$, translates the abstract representation to a process model and is inherited from conventional process discovery.

In the remainder, given an arbitrary \textit{data structure type} $\type$, we let $\typedDataStructureUniverse$ denote the universe of data structures of type $\type$.
A data type $\type$ might refer to an array or a (collection of) hash table(s), yet it might also refer to some implementation of a stream-based frequent-item approximation algorithm such as Lossy Counting~\cite{manku_2002_lossy}.
We assume any $\typedDataStructure \in \typedDataStructureUniverse$ to use finite memory.

\begin{definition}[Data Structure Update Function]
	\label{def:event_stream_data_structure}
	Let $\activity$ be a set of activities and let $\caseUniv$ denote the set of all possible case identifiers.
	We define a data structure update function $\dataStructureOnStreamFunction$ as:
\begin{equation}
\dataStructureOnStreamFunction \colon \typedDataStructureUniverse \times \caseAct \to \typedDataStructureUniverse 
\end{equation}
\end{definition}

The data structure update function $\dataStructureOnStreamFunction$ allows us to update a given data structure $\typedDataStructure \in \typedDataStructureUniverse$ based on any newly arrived event. 
In practice the function typically consists of two components.
One component keeps track of the cases that were already active before and maps them in some way to a second (collection of) data structure(s).
Such second component allows us to construct the abstract representation.
Thus, when abstracting this mechanism, given some event stream based data structure, we need a mechanism to translate the data structure, i.e. the range of $\dataStructureOnStreamFunction$, to an abstract representation.
\begin{definition}[Abstraction Function - Data Structure]
	\label{def:abstraction_funtion_data_structure}
	An abstraction function $\abstractionFunctionDataStrcuture$ is a function mapping a data structure of type $\type$ to an abstract representation of type $\typeAbstr$.
\begin{equation}
\abstractionFunctionDataStrcuture \colon \typedDataStructureUniverse \to \abstrTypedUniv
\end{equation}
\end{definition}
Ideally, translating the data structure is computationally inexpensive.
However, in some cases translating the data structure to the intended abstract representation might be expensive.
This is acceptable, as long as we (re)-compute the abstraction in a periodic fashion or at the user's request.

Assume that we have seen $i \geq 0$ events on an event stream $\stream$ and let $\typedDataStructureSub{i} \in \typedDataStructureUniverse$ denote the data structure that approximates the behaviour in the event stream $\stream$ after receiving $i$ events.
When new event $(\caseIdent,a) \in \caseAct$ arrives, we are able to discover a new process model $\model_{i+1}$ by applying $(\processDiscAlgoAbstr \circ \abstractionFunctionDataStrcuture \circ \dataStructureOnStreamFunction)(\typedDataStructureSub{i}, \caseIdent,a)$.
In practice, $\dataStructureOnStreamFunction$ is applied continuously and whenever, after receiving a new $i^{th}$ event, we are interested in finding a process model we apply $(\processDiscAlgoAbstr \circ \abstractionFunctionDataStrcuture)(\typedDataStructureSub{i})$ to obtain the process model.

The main challenge in instantiating the framework is designing a data structure $\typedDataStructure \in \typedDataStructureUniverse$ that allows us to approximate an abstract representation together with accompanying $\dataStructureOnStreamFunction$ and $\abstractionFunctionDataStrcuture$ functions.

\section{Instantiating S-BAR}
\label{sec:instantiations}
In this section, we show the applicability of the S-BAR framework by presenting several instantiations for different existing process discovery algorithms.
A large class of algorithms, e.g. the $\alpha$-Miner~\cite{aalst_2004_alpha}, the Heuristics Miner~\cite{weijters_2003_hm,weijters_2011_fhm} and the Inductive Miner~\cite{leemans_2013_inductive_miner}, is based on the directly follows abstraction.% or some derivative thereof.
Therefore, we first present how to compute this abstraction.
Subsequently we highlight, for each algorithm using the directly follows abstraction as a basis, the main changes and/or extensions that need to be applied w.r.t. the basic scheme.
To illustrate the generality of the architecture, we also show a completely different class of discovery approaches, i.e. region-based techniques~\cite{aalst_2010_state_based_region, werf_2009_ilp}.
These techniques work fundamentally different compared to the aforementioned class of algorithms and use different abstract representations.

\subsection{Directly Follows Abstraction}
\label{sec:direclty_follows_abstraction}
The \textit{directly follows abstraction} describes pairs of activities $(a,b)$, written as $a > b$, if there exists some sequence $\sigma \in L$ of the form $\sigma = \sigma' \cdot \langle a,b \rangle \cdot \sigma''$.
To approximate the relation, we let data structure $\typedDataStructure \in \typedDataStructureUniverse$ consist of two internal data structures $\mathcal{D^C}$ and $\mathcal{D^A}$.
Within $\mathcal{D^C}$ we store (case,activity)-pairs, i.e. $(c,a) \in C \times A$, that represent the last activity $a$ seen for case $c$.
Within $\mathcal{D^A}$ we store (activity, activity)-pairs $(a,a') \in A \times A$, where $(a,a') \in \mathcal{D^A} \Leftrightarrow a > a'$.
The basic scheme works as follows.
When a new event $(c,a)$ arrives, we check whether $\mathcal{D^C}$ already contains some pair $(c, a')$.
If so, we add $(a',a)$ to $\mathcal{D^A}$, remove $(c,a')$ from $\mathcal{D^C}$ and add $(c,a)$ to $\mathcal{D^C}$.
If not, we just add $(c,a)$ to $\mathcal{D^C}$.
$\mathcal{D^A}$ represents the directly follows abstraction by means of a collection of pairs, thus, function $\abstractionFunctionDataStrcuture$ consists of translating $\mathcal{D^A}$ to the appropriate underlying data type used by the discovery algorithm of choice.

As an example consider \autoref{alg:stream_space_saving} and \autoref{alg:stream_lossy_counting} describing a design of $\mathcal{D^C}$ based on the SpaceSaving algorithm~\cite{metwally_2005_space_saving} and Lossy Counting~\cite{manku_2002_lossy} respectively.
\begin{figure}[tb]
	\begin{minipage}[t]{0.5\textwidth}
		\vspace{0pt}
		\setlength{\algoheightrule}{0pt}
		\begin{algorithm}[H]		
			\scriptsize
			\caption{$\mathcal{D^C}$ (Space Saving)\label{alg:stream_space_saving}}
			\SetKwInOut{Input}{input}\SetKwInOut{Output}{output}
			\Input{$k \in \integersPos$, $\mathcal{S} \in (C \times A)^*$, $\mathcal{D^A}$} 
			%Output{$C \subseteq V$, representing the constraint body of the target ILP}
			\Begin{
				\nl $X \gets \emptyset$; $i \gets 0$\; 
				\nl \While{$\texttt{true}$} {
					\nl $i \gets i + 1$\;
					\nl $(c,a) \leftarrow\ \mathcal{S}\textit{(i)}$\;
					\nl	\If{$\exists_{(c',a') \in X}(c' = c)$}{
						\nl	$v_{c} \gets v_{c} + 1$\;
						\nl	$\mathcal{D^A} \uplus \{(a',a)\}$\;
						\nl	$X \gets (X \cup \{(c,a)\}) \setminus \{(c,a')\}$\;
					}
					\nl	\ElseIf{$|X| < k$}{
						\nl	$X \gets X \cup \{(c,a)\}$\;
						\nl	$v_{c} \gets 1$\;
					}
					\nl	\Else{
						\nl	$(c',a') \gets \underset{{(c',a') \in X}}{\arg\min} (v_{c'})$\;
						\nl	$\texttt{v}_{c} \gets \texttt{v}_{c'} + 1$\;
						\nl	$X \gets (X \cup \{(c,a)\}) \setminus \{(c',a')\}$\;
					}
				}
			}
		\end{algorithm}
	\end{minipage}%
	\begin{minipage}[t]{0.5\textwidth}
		\vspace{0pt}
		\setlength{\algoheightrule}{0pt}
		\begin{algorithm}[H]
			\scriptsize
			\caption{$\mathcal{D^C}$ (Lossy)\label{alg:stream_lossy_counting}}
			\SetKwInOut{Input}{input}\SetKwInOut{Output}{output}
			\Input{$k \in \integersPos$, $\mathcal{S} \in (C \times A)^*$, $\mathcal{D^A}$} 
			%Output{$C \subseteq V$, representing the constraint body of the target ILP}
			\Begin{
				\nl $i,\Delta \gets 0$; $X \gets \emptyset$\;
				\nl \While{$\texttt{true}$} {
					\nl $i \gets i + 1$\;
					\nl $(c,a) \leftarrow\ \mathcal{S}\textit{(i)}$\;
					\nl	\If{$\exists_{(c',a') \in X}(c' = c)$}{
						\nl	$v_c \gets v_c + 1$\;
						\nl	$\mathcal{D^A} \uplus \{(a',a)\}$\;
						\nl	$X \gets (X \cup \{(c,a)\}) \setminus \{(c,a')\}$\;
					}
					\nl	\Else{
						\nl	$X \gets X \cup \{(c,a)\}$\;
						\nl	$\texttt{v}_{c} \gets \Delta$\;
					}
					\nl \If{$\lfloor i / k \rfloor \neq \Delta$}{
						\nl \ForEach{$(c', a') \in X$}{
							\nl \If{$v_{c'} \leq \Delta$}{
								\nl $X \gets X \setminus (c',a')$\;
							}
						}
						\nl $\Delta \gets \lfloor i / k \rfloor$\;
					}
				}			
			}			
		\end{algorithm}
	\end{minipage}
\end{figure}
Both algorithms have three inputs, i.e. a maximum size $k \in \integersPos$, an event stream $\stream \in (C \times A)^*$ and a finite memory data structure implementing $\mathcal{D^A}$.
The algorithms maintain a set of (case,activity)-pairs $X$, initialized to $\emptyset$ (line~1).
For each case $c$ present in $X$ an associated counter $v_c$ is maintained which is used for memory management.
When a new event $(c,a)$ appears on the event stream, the algorithms check whether some pair $(c', a')$ s.t. $c = c'$ is stored in $X$ (line~5).
If this is the case, $c$'s counter is increased, $(a',a)$ is added to data structure $\mathcal{D^A}$ and $(c, a')$ is replaced by $(c,a)$ in $X$ (lines~6-8).
The algorithms differ in the way they process events $(c,a)$ for which $\nexists_{(c',a') \in X}(c' = c)$.
The Space Saving based algorithm (\autoref{alg:stream_space_saving}) either adds the element to $X$ if $|X| < k$ or replaces pair $(c',a') \in X$ with the lowest corresponding counter ($v_{c'}$) value (\autoref{alg:stream_space_saving}, lines~9-15).
The Lossy Counting based algorithm cleans up its $X$-set after each block of $k$ consecutive events and removes all those entries that have a counter value lower than variable $\Delta$ (lines~9-16).

Both algorithms insert a new element in data structure $\mathcal{D^A}$ in line~7.
Conceptually, the algorithms generate a stream of (activity, activity)-pairs.
Hence, in \autoref{alg:frequent} we present a basic design for $\mathcal{D^A}$ based on the Frequent Algorithm~\cite{DBLP:conf/esa/DemaineLM02,DBLP:journals/tods/KarpSP03} which uses an activity pair stream $\mathcal{S}_A \in (A \times A)^*$ as an input.
Thus, $\mathcal{D^A} \uplus \{(a',a)\}$ in line~7 of Algorithms~\ref{alg:stream_space_saving}~and~\ref{alg:stream_lossy_counting} represents adding pair $(a,a')$ at the end of stream $\mathcal{S}_A$.

\begin{wrapfigure}{I}{0.5\textwidth}
	\begin{minipage}{0.5\textwidth}
	\setlength{\algoheightrule}{0pt}
	\begin{algorithm}[H]
		\scriptsize
		\caption{$\mathcal{D^A}$ (Frequent)\label{alg:frequent}}
		\SetKwInOut{Input}{input}\SetKwInOut{Output}{output}
		\Input{$k \in \integersPos$, $\mathcal{S}_A \in (A \times A)^*$} 
		%Output{$C \subseteq V$, representing the constraint body of the target ILP}
		\Begin{
			\nl $X \gets \emptyset$, $i \gets 0$\;
			\nl \While{\texttt{true}} {
				\nl $i \gets i + 1$\;
				\nl $(a,a') \leftarrow\ \mathcal{S}_A\textit{(i)}$\;
				\nl	\If{$(a,a') \in X$}{
					\nl	$v_{(a,a')} \gets v_{(a,a')} + 1$\;
				}
				\nl	\ElseIf{$|X| < k$}{
					\nl	$X \gets X \cup \{(a,a')\}$\;
					\nl	$\texttt{v}_{(a,a')} \gets 1$\;
				}
				\nl	\Else{
					\nl \ForEach{$(x,y) \in X$}{
						\nl	$\texttt{v}_{(x,y)} \gets \texttt{v}_{(x,y)} - 1$\;
						\nl \If{$\texttt{v}_{(x,y)} = 0$}{
							\nl $X \gets X \setminus \{(x,y)\}$\;
						}
					}
				}
			}
		}
	\end{algorithm}
	\end{minipage}
\end{wrapfigure}
The algorithm stores pairs of activities in its internal set $X$.
Whenever a new pair $(a,a')$ arrives, the algorithm checks if it is already present in $X$, if so, it updates the corresponding counter $v_{(a,a')}$.
If the pair is not yet present in $X$, the size of $X$ is evaluated.
If $|X| < k$ the new pair is added to $X$ and a new counter is created for the pair.
If $|X| \geq k$ the new pair is not added, moreover, each counter is decreased by one and if a counter gets value 0 the corresponding pair is removed. 

The general mechanism of \autoref{alg:frequent} is very similar to \autoref{alg:stream_space_saving}.
The main difference consists of how to update $X$ when $|X| \geq k$.
All three algorithms use a parameter $k$ which, in a way, represents the (maximum) size of $X$.
Hence, when we write $|\mathcal{D^C}|$, $|\mathcal{D^A}|$ respectively, we implicitly refer to the value of $k$.
It should be clear that we are also able to implement $\mathcal{D^C}$ based on the Frequent Algorithm, i.e. we just adopt a different updating mechanism for $X$.
Likewise we are also able to design $\mathcal{D^A}$ based on the Space Saving/Lossy Counting algorithm.
Moreover, for $\mathcal{D^C}$ we are able to use other types of stream-aware data structures, i.e. techniques adopting a different scheme to ensure finite memory.
Examples of such types of techniques are Reservoir Sampling~\cite{aggarwal_2006_biased_rs}, and/or Decay Based Schemes~\cite{cormode_2009_forward_decay}.
In the next sections we briefly explain how the $\alpha$-Miner, Heuristics Miner and Inductive Miner use the directly follows abstraction and what changes to the base scheme must be applied in order to adopt them in a streaming setting.

\subsubsection{The $\alpha$-Miner}
\label{sec:alpha_miner}
The $\alpha$-Miner~\cite{aalst_2004_alpha} transforms the directly follows abstraction into a Petri net.
When adopting the $\alpha$-Miner to an event stream context, we directly adopt the scheme described in the previous section.
However, the algorithm explicitly needs a set of \textit{start-} and \textit{end activities}.

Approximating the start activities seems rather simple, i.e. whenever we receive a new case, the corresponding activity represents a start activity.
However, given that we at some point remove (case,activity)-pairs from $\mathcal{D^C}$, we might designate some activities falsely as start activities, i.e. a new case may in fact refer to a previously removed case.
Approximating the end activities is more complex, as we are often not aware when a case terminates.
A potential solution is to apply a \textit{warm-up} period in which we try to observe cases that seem to be terminated, e.g. by identifying cases that have long periods of inactivity or by assuming that cases that are dropped out of $\mathcal{D^C}$ are terminated.
However, since we approximate case termination, using this approach may lead to falsely select certain activities as end activities.

We can also deduce start- and end activities from the directly follows abstraction.
A start activity is an $a \in A$ with $\nexists_{a' \in A}(a' \neq a \mid a' > a)$ and an end activity is an $a \in A$ with $\nexists_{a' \in A}(a' \neq a \mid a > a')$.
This works if these activities are only executed once at the beginning, respectively the end, of the process.
In case of loops or multiple executions of start/end activities within the process, we potentially falsely neglect certain activities as being either start and/or end activities.
In \autoref{sec:init_termination}, we discuss this problem in depth.

\subsubsection{The Heuristics Miner}
\label{sec:heuristics_miner}
The Heuristics Miner~\cite{weijters_2003_hm, weijters_2006_hm_techrep,weijters_2011_fhm} is designed to cope with noise in event logs.
To do this, it effectively counts the number of occurrences of activities, as well as the $>$-relation.
Based on the directly follows abstraction it computes a derived metric $a \Rightarrow b = \frac{|a>b| - |b>a|}{|a > b| + |b > a| + 1}$ that describes the relative causality between two tasks $a$ and $b$ ($|a > b|$ denotes the number of occurrences of $a >  b$).
The basic scheme presented in \autoref{sec:direclty_follows_abstraction} suffices for computing $a \Rightarrow b$, as long as $\mathcal{D^A}$ explicitly tracks, or, approximates, the frequencies of its elements (in the scheme this is achieved by the internal counters).

\subsubsection{The Inductive Miner}
\label{sec:inductive_miner}
The Inductive Miner~\cite{leemans_2013_inductive_miner}, like the $\alpha$-Miner, uses the directly follows abstraction and start and end activities.
It tries to find patterns within the directly follows abstraction that indicate certain behaviour, e.g. parallelism.
Using these patterns it splits the event log into several smaller logs and repeats the procedure.
Due to its iterative nature, the Inductive Miner guarantees to find \textit{sound workflow nets}~\cite{aalst_1998_workflow_nets}.
The Inductive Miner has also been extended to handle noise and/or infrequent behaviour~\cite{leemans_2013_inductive_infrequent}.
This requires, like the Heuristics Miner, to count the $>$-relation.
In~\cite{leemans_2015_scalable_guarantee}, a version of the Inductive Miner is presented in which the inductive steps are directly performed on the directly follows abstraction.
In context of event streams this is the most adequate version to use as we only need to maintain a (counted) directly follows abstraction.

\subsection{Region Theory}
Several process discovery algorithms~\cite{aalst_2010_state_based_region,bergenthum_2007_process_mining_regions,carmona_2014_abstract_domains,werf_2009_ilp,zelst_2015_ilp} are based on \textit{region theory} which solve the Petri net synthesis problem~\cite{badouel_2015_petri_net_synthesis}.
Classical region theory techniques ensure strict formal properties for the resulting process models.
Process discovery algorithms based on region theory relax these properties.
We identify two different region theory approaches, i.e. \textit{language-based} and \textit{state-based} region theory, which use different forms of abstract representations.

\subsubsection{Language-Based Approaches}
Algorithms based on \textit{language-based} region theory~\cite{bergenthum_2007_process_mining_regions,werf_2009_ilp} rely on a \textit{prefix-closure} of the input event log, i.e. the set of all prefixes of all traces.
It is trivial to adapt the scheme presented to compute the directly follows abstraction (\autoref{sec:direclty_follows_abstraction}) to prefix-closures.
In stead of storing (case,activity)-pairs in $\mathcal{D^C}$, we store pairs $(c,\sigma) \in C \times A^*$.
We additionally use a data structure $\mathcal{D}^{pc}$ which approximates the prefix-closure. 
Whenever we receive an event $(c,a)$, we look for a pair $(c,\sigma) \in \mathcal{D^C}$.
If such pair exist we subsequently add $\sigma' = \sigma \cdot \langle a \rangle$ to $\mathcal{D}^{pc}$ and update $(c,\sigma)$ to $(c,\sigma')$.
If there is no such pair $(c, \sigma)$, we add $\epsilon$ and $\langle a \rangle$ to $\mathcal{D}^{pc}$ and $(c, \langle a \rangle)$ to $\mathcal{D^C}$.
In case of~\cite{werf_2009_ilp}, which uses Integer Linear Programming where (an abstraction of) the prefix-closure forms the constraint body, we simply store the constraints in $\mathcal{D}^{pc}$, rather than the prefix-closure.

\subsubsection{State-Based Approaches}
Within process discovery based on state-based regions~\cite{aalst_2010_state_based_region}, a transition system is constructed based on \textit{a view} of a trace.
Examples of a view are the complete prefix of the trace, the multiset projection of the prefix, etc.
The \textit{future of a trace} can be used as well, i.e. given an event within a trace, the future of the event are all events happening after the event.
However, future-based views are not applicable in an event stream setting, as the future is unknown.

\begin{figure}[b]
	\centering	
	\subfloat[Multiset abstraction (unbounded).]{
		\resizebox{0.4\textwidth}{!}{
			\begin{tikzpicture}[->,>=stealth',shorten >=1pt,auto,node distance=0.75cm,main node/.style={circle,fill=white,draw,minimum size=0pt, inner sep=0pt}]		
			\node[main node] (eps) [label=above:{\scriptsize $s_0: [\ ]$}]{};
			\node[main node] (n1) [label=right:{\scriptsize $s_1: [a]$}, below = of eps]{};
			\node[main node] (n2) [label=right:{\scriptsize $s_2: [a,b]$}, below right = of n1]{};
			\node[main node] (n3) [label=left:{\scriptsize $s_3: [a,c]$}, below left = of n1]{};
			\node[main node] (n4) [label=right:{\scriptsize $s_4: [a,b,c]$}, below left = of n2, opacity=0.5]{};
			\node[main node] (n5) [label=right:{\scriptsize $s_5: [a,b,c,d]$}, below = of n4]{};
			
			\path[->]
			(eps) edge node [] {\scriptsize $a$} (n1)
			(n1) edge node [right] {\scriptsize $b$} (n2)
			(n1) edge node [left, pos=0.25] {\scriptsize $c$} (n3)
			(n2) edge node [right]{\scriptsize $c$} (n4)
			(n3) edge node [left, pos=0.65] {\scriptsize $b$} (n4)
			(n4) edge node [right] {\scriptsize $d$} (n5);		
			\end{tikzpicture}
			\label{fig:example_ts_multiset}
		}
	}
	\subfloat[Set abstraction (max. set size $1$).]{
		\resizebox{0.4\textwidth}{!}{
			\begin{tikzpicture}[->,>=stealth',shorten >=1pt,auto,node distance=1cm,main node/.style={circle,fill=white,draw,minimum size=0pt, inner sep=0pt}]		
			\node[main node] (eps) [label=above:{\scriptsize $s_0: \emptyset$}]{};
			\node[main node] (n1) [label=right:{\scriptsize $s_1: \{a\}$}, below = of eps]{};
			\node[main node] (n2) [label=right:{\scriptsize $s_2: \{b\}$}, below right = of n1]{};
			\node[main node] (n3) [label=left:{\scriptsize $s_3: \{c\}$}, below left = of n1]{};
			\node[main node] (n4) [label=right:{\scriptsize $s_4: \{d\}$}, below left = of n2, opacity=0.5]{};
			
			\path[->]
			(eps) edge node [right] {\scriptsize $a$} (n1)
			(n1) edge node [right] {\scriptsize $b$} (n2)
			(n1) edge node [left, pos=0.25] {\scriptsize $c$} (n3)
			(n2) edge [bend left = 15] node []{\scriptsize $c$} (n3)
			(n3) edge [bend left = 15] node [] {\scriptsize $b$} (n2)
			(n2) edge node [right]{\scriptsize $d$} (n4)
			(n3) edge node [left] {\scriptsize $d$} (n4);
			\end{tikzpicture}
			\label{fig:example_ts_set}
		}
}
		\caption{Example transition systems based on $L=[\langle a,b,c,d \rangle, \langle a,c,b,d \rangle]$.}
		\label{fig:example_ts}
\end{figure}
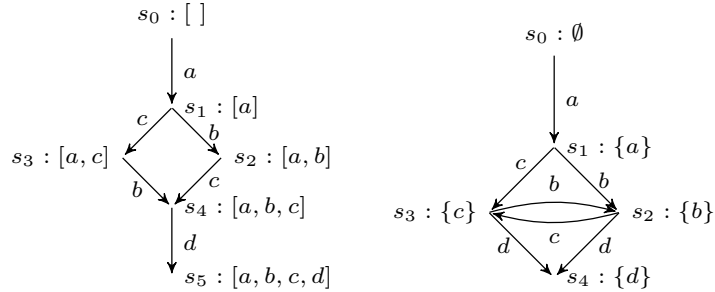

As an example of a transition system based on a simple event log $L=[\langle a,b,c,d \rangle, \langle a,c,b,d \rangle]$, consider \autoref{fig:example_ts}.
In \autoref{fig:example_ts_multiset} states are represented by a multiset view of the prefixes of the traces, i.e. the state is determined by the multiset of activities seen before.
Activities make up the transitions within the system, i.e. the first activity in both traces is $a$, thus the empty multiset is connected to multiset $[a]$ by means of a transition labelled $a$.
In \autoref{fig:example_ts_multiset} we do not limit the maximum size of the multisets.
\autoref{fig:example_ts_set} shows a set view of the traces with a maximum set size of $1$.
Again the empty set is connected with set $\{a\}$ by means of a transition labelled $a$.
For trace $\langle a,b,c,d \rangle$ for example, the second activity is a $b$ and thus state $\{a\}$ has an outgoing transition labelled $b$ to state $\{b\}$.
This is the case, i.e. a connection to state $\{b\}$ rather than $\{a,b\}$, due to the size restriction of size $1$.

Consider the following scheme, similar to the scheme presented in \autoref{sec:direclty_follows_abstraction}.
Given a view type $V$, e.g. a set view, we design $\mathcal{D^C}$ to maintain pairs $(c, v_{c})$, s.t. $v_{c}$ is the last view constructed for case $c$.
Moreover, we maintain a collection of views $\mathcal{D}^V$.
Updating $\mathcal{D}^V$ is straightforward.
Given new event $(c,a)$, based on $v_{c}$ we compute some new view $v'_{c}$, add it to $\mathcal{D}^V$ and update $(c, v_{c})$ to $(c, v'_{c})$ in $\mathcal{D^C}$, e.g. updating the size-1 set view means that the new view based on new event $(c,a)$ is simply the set $\{a\}$.
However, just maintaining size-1 sets in $\mathcal{D}^V$ does not suffice as the relations between those sets, i.e. the transitions in the transition system, are not present in $\mathcal{D}^V$.

The problem is fixed by maintaining the transition system in memory, rather than $\mathcal{D}^V$, and updating it directly when we receive new events.
Given some latest view $v_{c}$ for case $c$, i.e. $(c, v_{c}) \in \mathcal{D^C}$, activity $a$ of new event $(c, a)$ represents the transition from $v_{c}$ to the newly derived $v'_{c}$.
Without a limit on the view-size, translating the transition system into a Petri net is rather slow.
Hence, in a streaming setting we limit the maximum size of the views.
This, in turn, causes some challenges w.r.t. $\mathcal{D^C}$ and translation function $\abstractionFunctionDataStrcuture$.
Consider the case where we maintain a multiset/set view of traces with some arbitrary finite capacity $k$.
Moreover, given $k = 2$, assume we receive event $(c, a)$ and $(c, \{a',a''\}) \in \mathcal{D^C}$.
The question is whether the new view for $c$ is $\{a, a'\}$ or $\{a,a''\}$?
Only if we store the last two events observed for $c$, in order, we are able to answer this question, i.e. if $(c, \langle a', a'' \rangle) \in \mathcal{D^C}$ we deduce the new view to be $\{a, a''\}$.
Finally note, that when we aim at removing paths from the transitions system, for example when we remove cases from $c$ from $\mathcal{D^C}$, we need to store the whole trace for $c$ in order to be able to reduce all states and transitions related to case $c$.

\section{Evaluation}
\label{sec:evaluation}
In this section we present an evaluation of several instantiations of the architecture.
We also consider performance aspects of the implementation.
All five algorithms, i.e. $\alpha$-Miner, Heuristics Miner, Inductive Miner, ILP (language based regions) and Transition System Miner (state based regions), have been implemented using the schemes presented in \autoref{sec:instantiations} in the ProM~\cite{dongen_2005_prom} framework (\url{http://www.promtools.org}).
ProM is the de facto standard academic tool-kit for process mining algorithms and is additionally used by practitioners in the field.
Some of the implementations are ported to RapidProM~\cite{DBLP:journals/corr/AalstBZ2017} (\url{http://www.rapidprom.org}), i.e. a plugin of RapidMiner (\url{http://www.rapidminer.com}), which allows for designing large-scale repetitive experiments by means of scientific workflows~\cite{bolt_2015_swf}.
Source code of the implementations is available via the \textit{Stream}-related packages within the ProM code base, i.e. \textit{StreamAbstractRepresentation},  \textit{StreamAlphaMiner}, \textit{StreamHeuristicsMiner}, \textit{StreamILPMiner}, \textit{StreamInductiveMiner} and \textit{StreamTransitionSystemsMiner} (code for a package X is located at \url{http://svn.win.tue.nl/repos/prom/Packages/X}).
Experiment results, event streams and generating process models used, are available at \url{https://github.com/s-j-v-zelst/research/releases/download/kais1/2016_kais1_experiments.tar.gz}.

\subsection{Structural Analysis}
As a first visual experiment we investigate the steady-state behaviour of the \textit{Inductive Miner}~\cite{leemans_2013_inductive_miner}.
For both $\mathcal{D^C}$ and $\mathcal{D^A}$ we use the Lossy Counting scheme (\autoref{sec:direclty_follows_abstraction}).
To create an event stream, we created a timed Coloured Petri Net~\cite{jensen_2009_cpn} in CPN-Tools~\cite{jensen_2007_cpn_tools} which simulates the BPMN model depicted in \autoref{fig:running_example_bpmn} and emits the corresponding events.
The event stream, and all other event streams used for experiments, are free of noise.
The model is able to simulate multiple cases being executed simultaneously.
The ProM streaming framework~\cite{zelst_2014_prom_streams, zelst_2015_stream_cpn} is used to generate an event stream out of the process model. 

\begin{figure}[t]
	\includegraphics[width=0.95\textwidth]{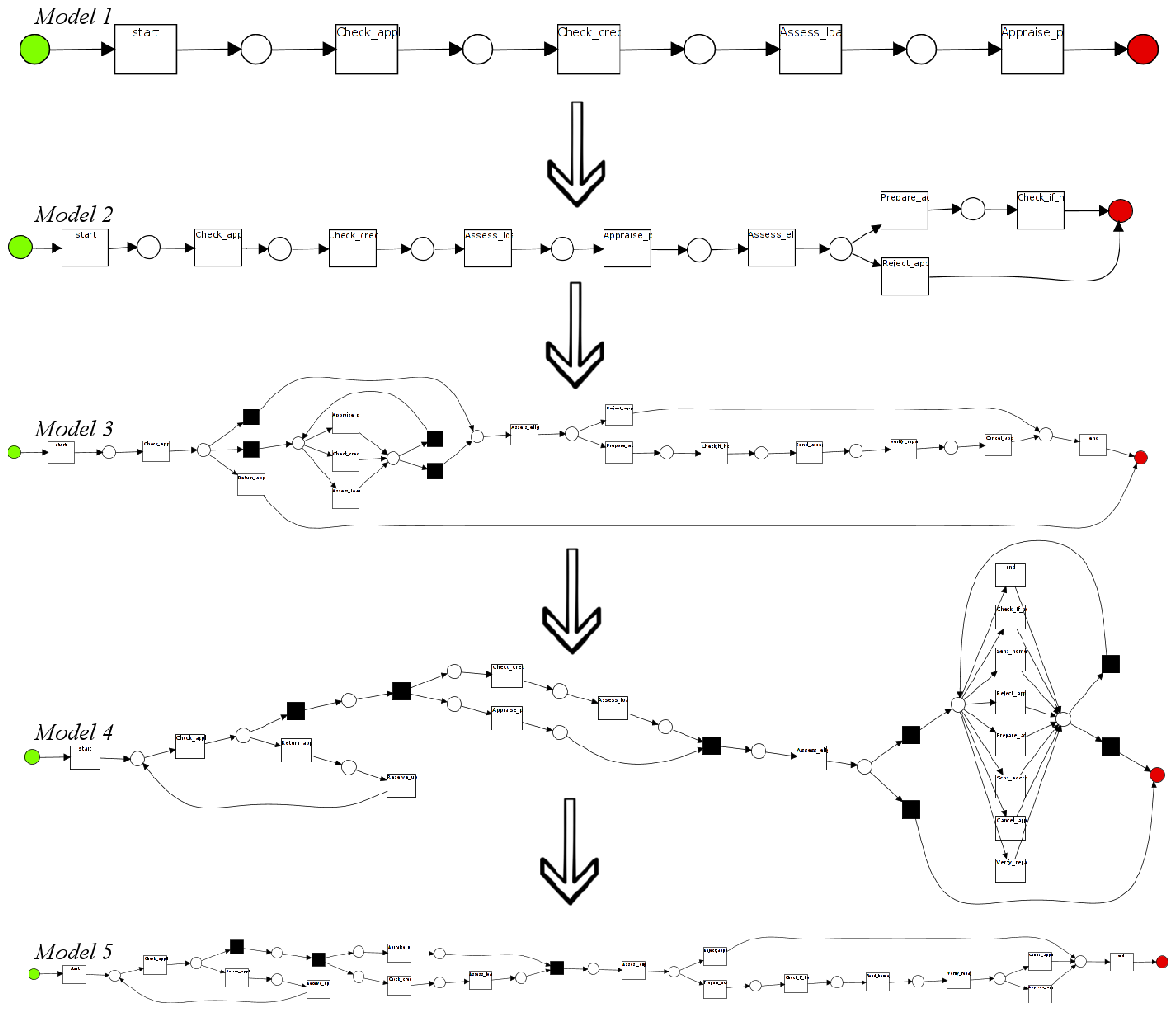}
	\caption{Visual results of applying the Inductive Miner on a stream.}
	\label{fig:demo}
\end{figure}

In \autoref{fig:demo} we show the behaviour of the Inductive Miner over time, configured with $|\mathcal{D^C}| = 75$, $|\mathcal{D^A}| = 75$, based on a random simulation of the CPN model.
Initially (Model 1) the Inductive Miner only observes a few directly follows relations, all executed in sequence.
After a while (Model 2) the Inductive Miner observes that there is a choice between \textit{Prepare acceptance pack} and \textit{Reject Application}.
In Model 3, the first signs of parallel behaviour of activities \textit{Appraise property}, \textit{Check credit history} and \textit{Assess loan risk} become apparent.
However, not enough behaviour is emitted onto the stream to effectively observe the parallel behaviour yet.
In Model 4, we identify a large block of activities within a choice construct.
Moreover, an invisible transition loops back into this block.
The Inductive Miner tends to show this type of behaviour given an incomplete directly follows abstraction.
Finally, after enough behaviour is emitted onto the stream, Model 5 shows a Petri net version of example process model of \autoref{fig:running_example_bpmn}.

\autoref{fig:demo} shows that the Inductive Miner is able to find the original model based on the event stream.
We now focus on comparing the Inductive Miner with other algorithms described in the paper.	
All discovery techniques discover a Petri net or some alternative process model that we can \textit{convert to} a Petri net.
The techniques however differ in terms of guarantees w.r.t. the resulting process model.
The Inductive Miner guarantees that the resulting Petri nets are \textit{sound}, whereas the ILP Miner and the Transition System Miner do not necessarily yield sound process models.
To perform a proper behavioural comparative analysis, the soundness property is often a prerequisite.
Hence, we perform a structural analysis of all the algorithms by measuring structural properties of the resulting Petri nets.

Using the off-line variant of each algorithm we first compute a reference Petri net.
We generated an event log $L$ which contains \textit{enough behaviour} such that the discovered Petri nets describe all behaviour of the BPMN model of \autoref{fig:running_example_bpmn}.
Based on the reference Petri net we create a 15-by-15 matrix in which each row/column corresponds to an activity in the BPMN model.
If, in the Petri net, two labelled transitions are connected by means of a place, the corresponding cells in the matrix get value $1$.
For example, given the first Petri net of \autoref{fig:demo}, the labels \textit{start} and \textit{Check\_application\_completeness} (in the figure this is ``Check\_appl'') are connected by means of a place.
Hence, the distance between the two labels is set to $1$ in the corresponding matrix.
If two transitions are not connected, the corresponding value is set to $0$.

Using an event stream based on the CPN-Model, after each newly received event, we use each algorithm to discover a Petri net.
For each Petri net we construct the 15-by-15 matrix.
We apply the same procedure as applied on the reference model.
However, if in a discovered Petri net a certain label is not present, we set all cells in the corresponding row/column to $-1$, e.g. in model 1 of \autoref{fig:demo} there is no transition labelled \textit{end}, thus the corresponding row and column consist of $-1$ values.
Given a matrix $M$ based on the streaming variant of an algorithm, we compute the distance to the reference matrix $M_R$ as: $d_{M, M_R} = \sqrt{\sum_{i,j \in \{1,2,...,15\}}((M(i,j) - M_R(i,j))^2}$.
For all algorithms, the internal data structures used where based on Lossy Counting, with size $100$.

\begin{figure}[tb]
	\centering
	\subfloat[Distances for $\alpha$ and IM]{
		\includegraphics[width=0.5\textwidth]{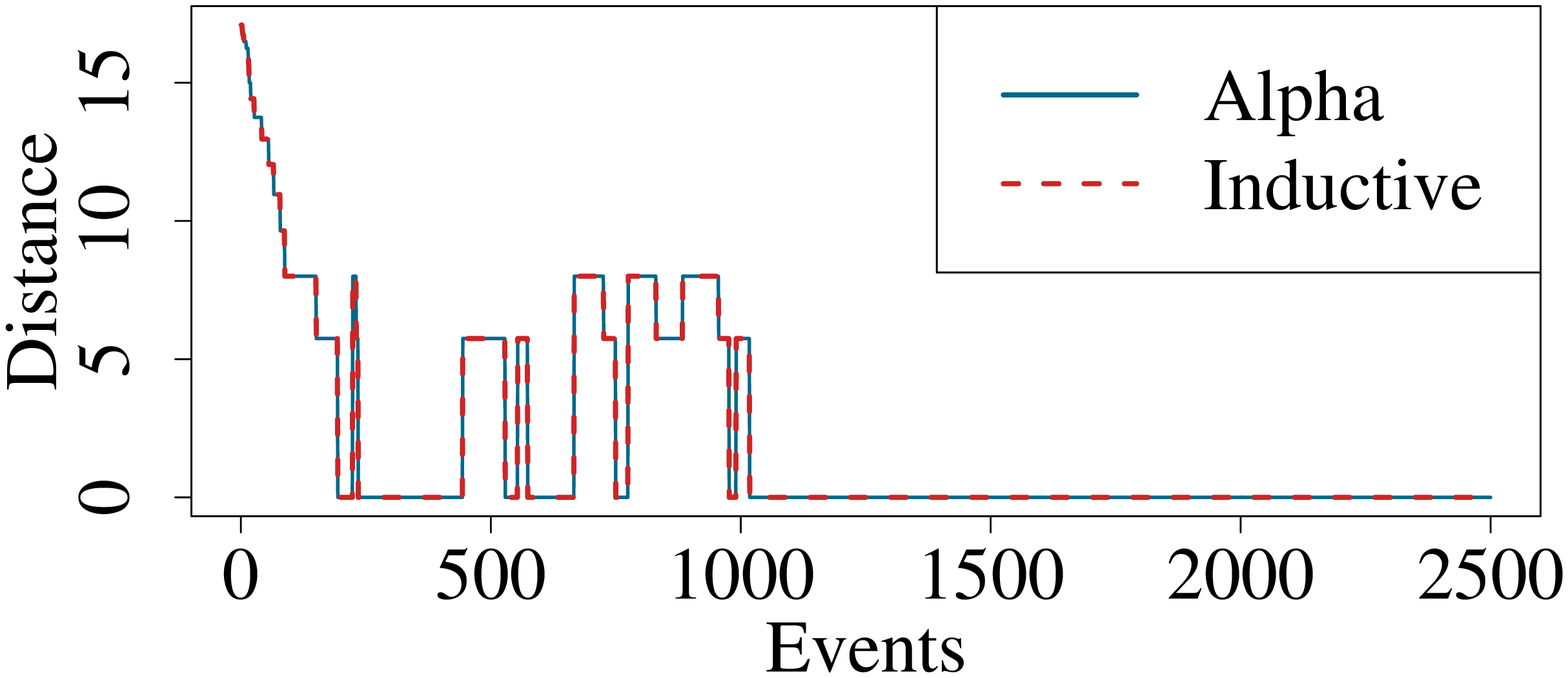}
		\label{fig:exp_0_eucl_alpha_ind}
	}
	\subfloat[Distances for TS, ILP and HM.]{
		\includegraphics[width=0.5\textwidth]{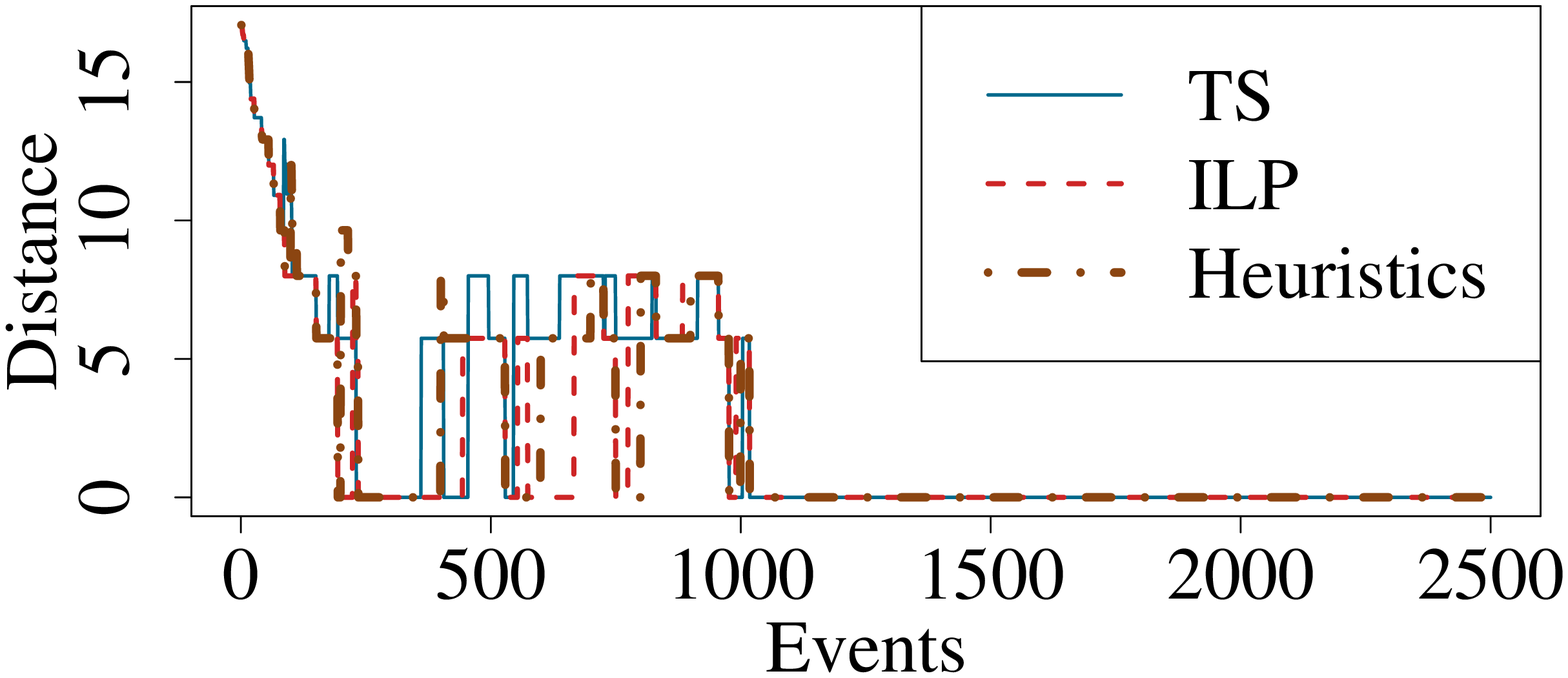}
		\label{fig:exp_0_eucl_ts_ilp_heur}
	}   
	\caption{Distance measurements for the $\alpha$-Miner, Inductive Miner (IM), ILP Miner (ILP), Transition Systems Miner (TS) and Heuristics Miner.}
	\label{fig:distance}
\end{figure}

Since the Inductive Miner and the $\alpha$-Miner are completely based on the same abstraction, we expect them to behave similar.
Hence, we plot their corresponding results together in \autoref{fig:exp_0_eucl_alpha_ind}.
Interestingly, the distance metric follows the same pattern for both algorithms.
Initially, there is a steep decline in the distance metric after which it becomes zero.
This means that the reference matrix equals the matrix based on the discovered Petri net.
The distance shows some peaks in the area between 400 until 1000 received events.
Analyzing the resulting Petri nets at these points in time showed that some activities where not present in the resulting Petri nets at those points.
The results for the Transition Systems Miner (TS), the ILP Miner and the Heuristics Miner are depicted in \autoref{fig:exp_0_eucl_ts_ilp_heur}.
We observe that the algorithms behave similar to the $\alpha$- and Inductive Miner, which intuitively makes sense as the algorithms all have the same data structure capacity.
However, the peeks in the distance metric occur at different locations.
For the Heuristics Miner this is explained by the fact that it takes frequency into account and thus uses the directly follows abstraction differently.
The Transition System Miner and the ILP Miner use different abstract representations, and have a different update mechanism than the directly follows abstraction, i.e. they always update their abstraction whereas the directly follows abstraction only updates if, for a given case, we already received a preceding activity.

\subsection{Behavioral Analysis}
Although the previous experiments provide interesting insights w.r.t. the functioning of the algorithms in a streaming setting, they only consider structural model quality.
A distance value of 0 in \autoref{fig:distance} indicates that the resulting model is very similar to the reference model.
It does not guarantee that the model is in fact equal, or, entails the same behaviour as the reference model.
Hence, in this section we focus on measuring quantifiable similarity in terms of \textit{behaviour}.
We use the Inductive Miner as it provides formal guarantees w.r.t. initialization and termination of the resulting process models.
This in particular is a requirement to measure behavioural similarity in a reliable manner.
We adapt the Inductive Miner to a streaming setting by instantiating the S-BAR framework, using the scheme described in \autoref{sec:direclty_follows_abstraction}, combined with the modifications described in \autoref{sec:inductive_miner}.
For finding start and end activities we traverse the directly follows abstraction and select activities that have no predecessor, or, successor, respectively.
We again use Lossy Counting~\cite{manku_2002_lossy} to implement both $\mathcal{D^C}$ and $\mathcal{D^A}$ (\autoref{alg:stream_lossy_counting}, \autoref{sec:direclty_follows_abstraction}).

\begin{figure}[tb]
	\centering
	\subfloat[$|\mathcal{D^C}| = 25$, $|\mathcal{D^A}| = 25$]{
		\includegraphics[width=0.5\textwidth]{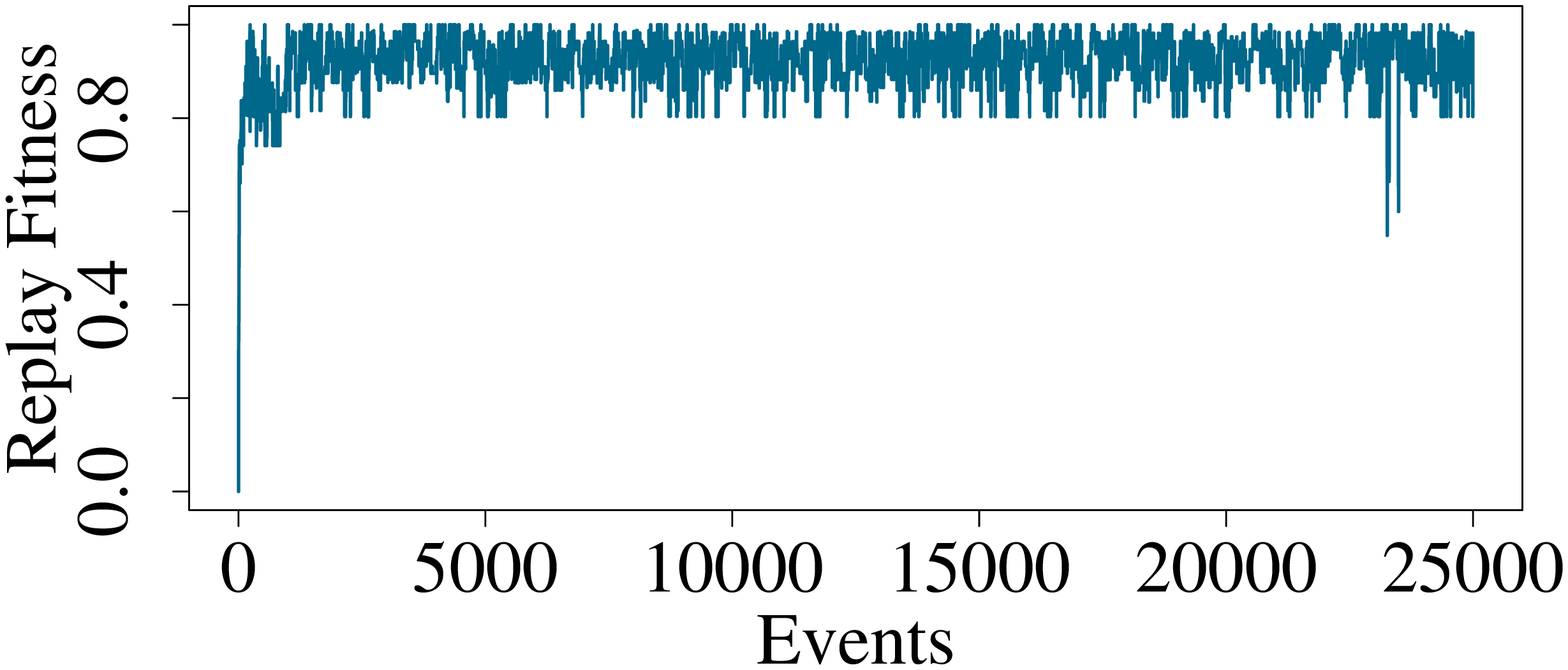}
		\label{fig:exp_1_im_fitness_25_25}
	}
	\subfloat[$|\mathcal{D^C}| = 75$, $|\mathcal{D^A}| = 75$]{
		\includegraphics[width=0.5\textwidth]{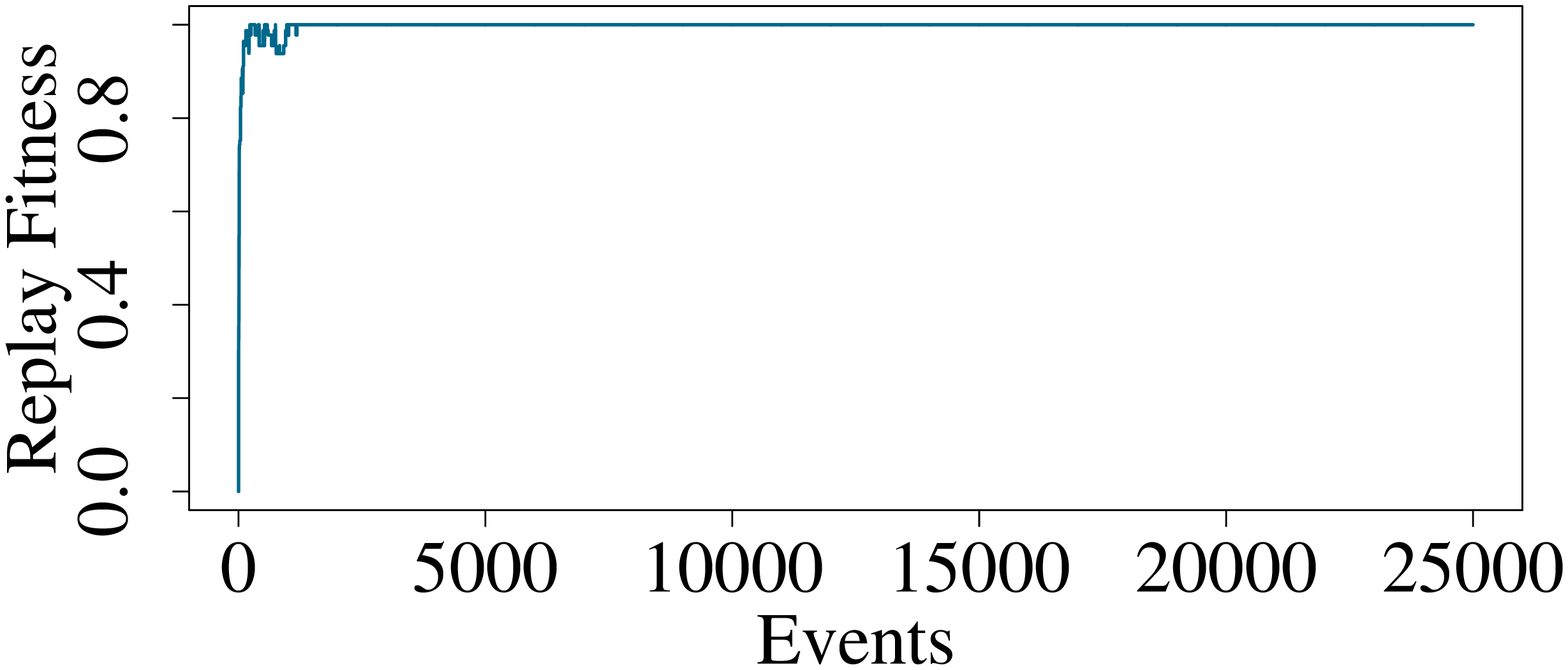}
		\label{fig:exp_1_im_fitness_75_75}
	}
	
	\subfloat[$|\mathcal{D^C}| = 25$, $|\mathcal{D^A}| = 25$]{
		\includegraphics[width=0.5\textwidth]{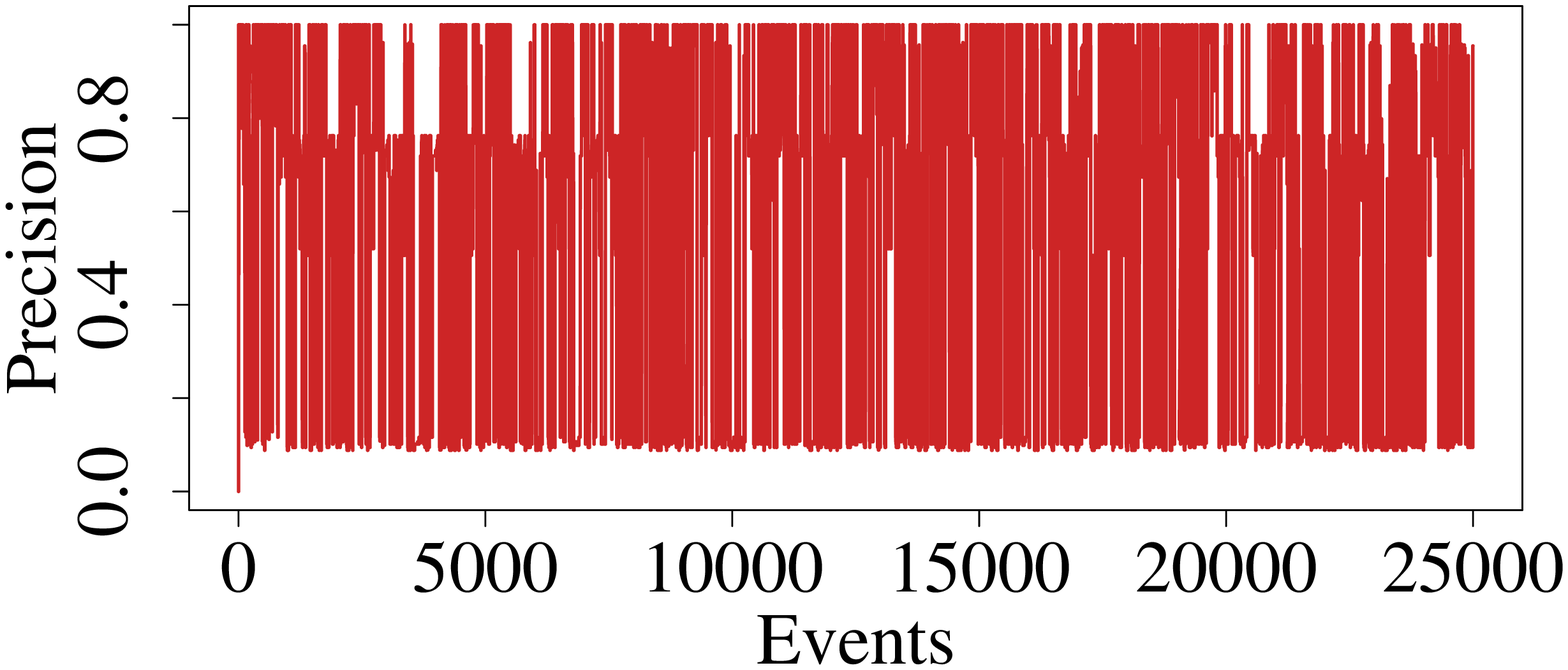}
		\label{fig:exp_1_im_precision_25_25}
	}
	\subfloat[$|\mathcal{D^C}| = 75$, $|\mathcal{D^A}| = 75$]{
		\includegraphics[width=0.5\textwidth]{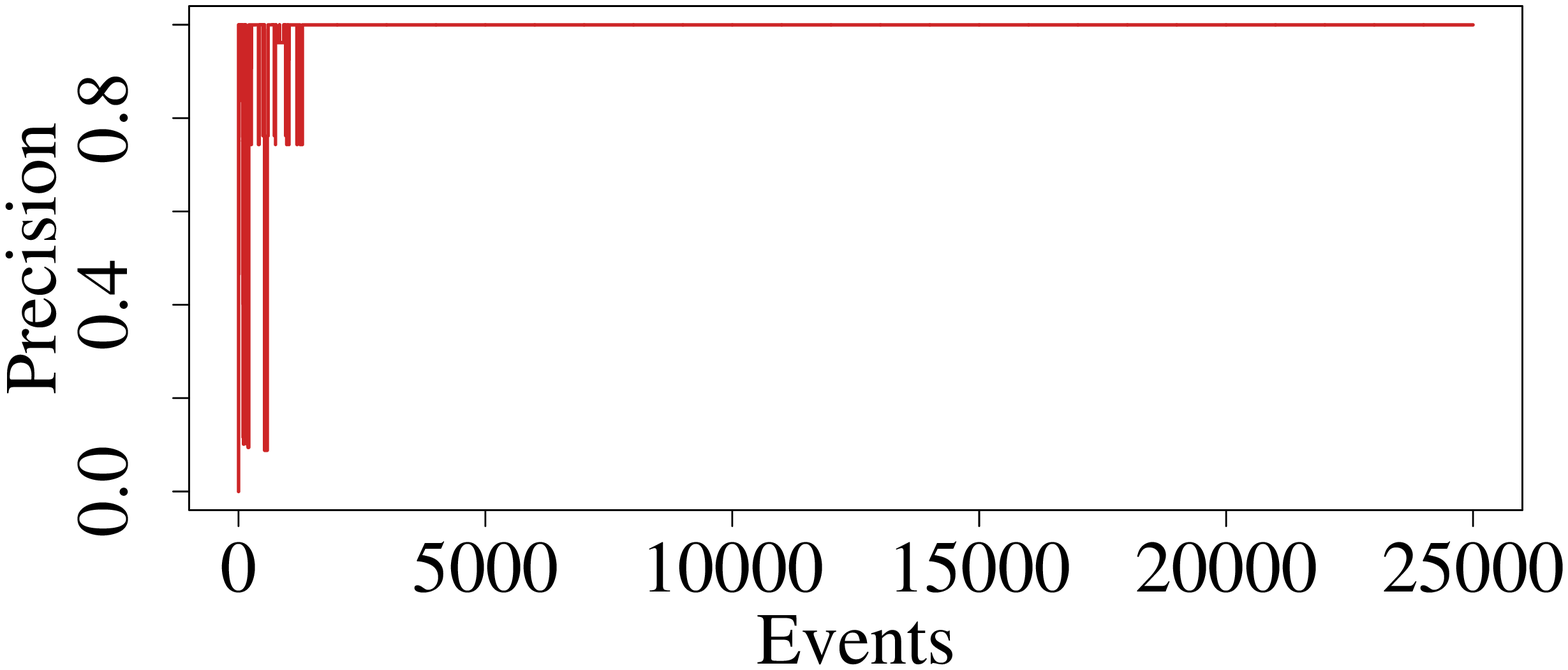}
		\label{fig:exp_1_im_precision_75_75}
	}
	\caption{Replay fitness and Precision measures based on applying the Stream Inductive Miner: Increasing memory helps to improve fitness and precision.}
	\label{fig:exp_1_im_fitness_precision_results}
\end{figure}

We assess under what conditions the Inductive Miner instantiation is able to discover a process model with the same behaviour as the BPMN model in \autoref{fig:running_example_bpmn}.
In the experiment, after each received event, we query the miner for its current result and compute replay fitness and precision measures based on a complete corresponding event log.
In \autoref{fig:exp_1_im_fitness_precision_results} the results are presented for varying capacity sizes of the underlying data structure (Lossy Counting).
%FITNESS

For the smallest data structure sizes, i.e. \autoref{fig:exp_1_im_fitness_25_25}, we identify that the replay fitness does not stabilize.
When the data structure size increases, i.e. \autoref{fig:exp_1_im_fitness_75_75}, we identify the replay fitness to reach a value of 1 rapidly.
The high variability in the precision measurements present in \autoref{fig:exp_1_im_precision_25_25} suggests that the algorithm is not capable of storing the complete directly follows abstraction.
As a result, the Inductive Miner tends to create flower-like patterns, thus greatly under-fitting the actual process.
The stable pattern present in \autoref{fig:exp_1_im_precision_75_75} suggest that the sizes used within the experiment are sufficient to store the complete directly follows abstraction.
Given that the generating process model is within the class of \textit{re-discoverable process models} of the Inductive Miner, both a replay fitness and a precision value of 1 indicates that the model is completely discovered by the algorithm.

\begin{figure}[tb]
	\centering
	\subfloat[$|\mathcal{D^C}| = 100$, $|\mathcal{D^A}| = 10,20...,50$]{
		\includegraphics[width=0.5\textwidth]{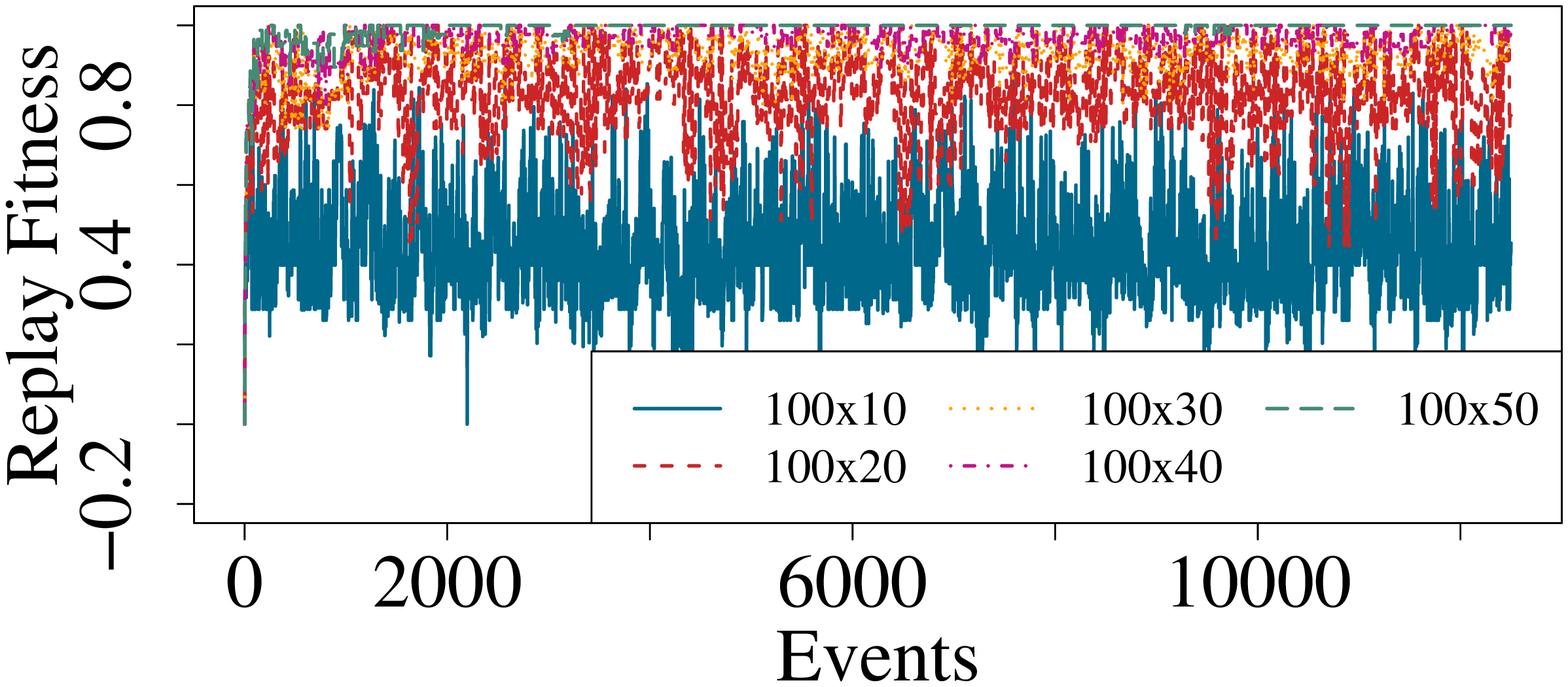}
		\label{fig:ex1_small_ca_100}
	}
	\subfloat[$|\mathcal{D^C}| = 10,20,...,50$, $|\mathcal{D^A}| = 100$]{
		\includegraphics[width=0.5\textwidth]{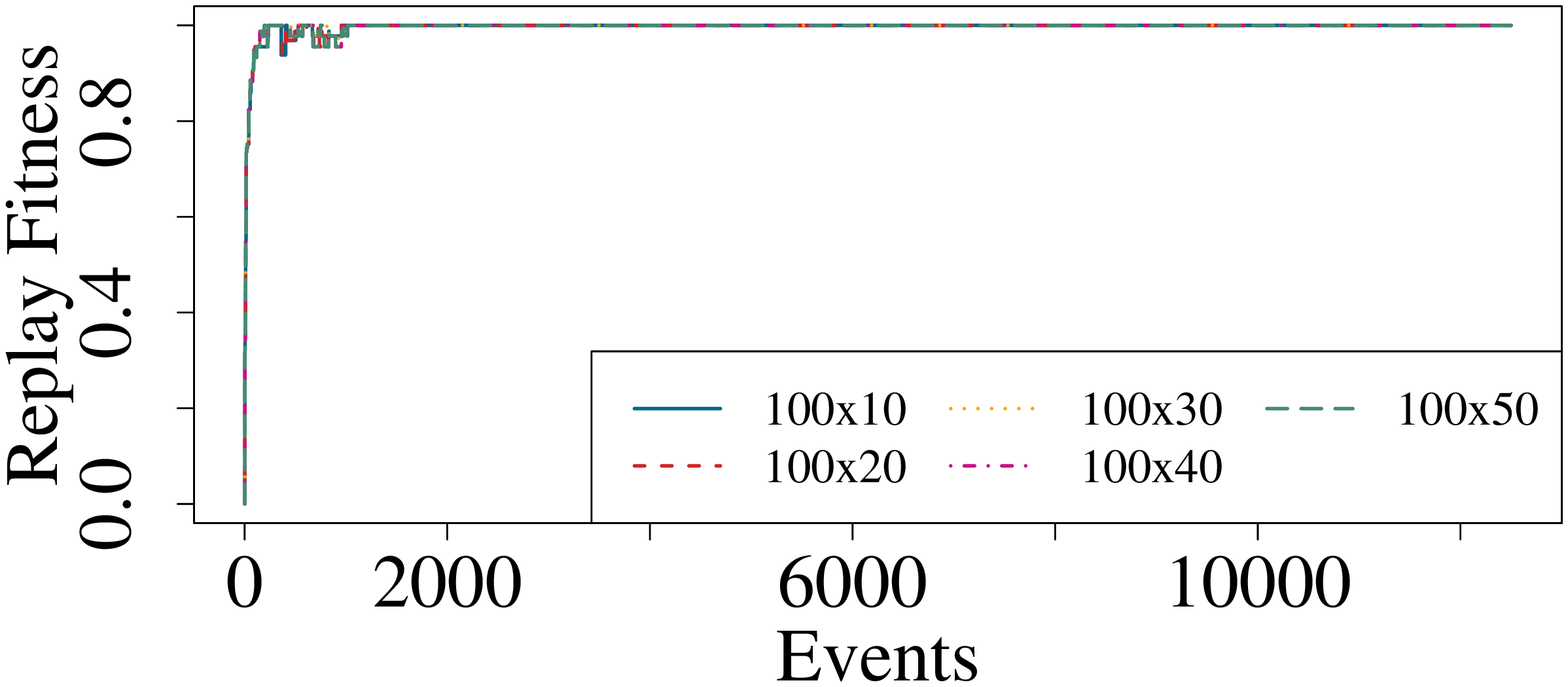}
		\label{fig:ex1_small_aa_100}
	}
	\caption{Replay Fitness measures for the Stream Inductive Miner.}
	\label{fig:exp_1_small_ds}
\end{figure}

In the previous experimental setting, we chose to use the same capacity for both $\mathcal{D^C}$ and $\mathcal{D^A}$.
Here we study the influence of the individual sizes of $\mathcal{D^C}$ and $\mathcal{D^A}$.
In \autoref{fig:exp_1_small_ds} we depict the results of two different experiments in which we fixed the size of one of the two data structures and varied the size of the other data structure.
\autoref{fig:ex1_small_ca_100} depicts the results for a fixed value $|\mathcal{D^C}| = 100$ and varying sizes $|\mathcal{D^A}| = 10,20,...,50$.
\autoref{fig:ex1_small_aa_100} depicts the results for a fixed value $|\mathcal{D^A}| = 100$ and varying sizes $|\mathcal{D^C}| = 10,20,...,50$.
As the results show, the lack of conversion to a replay fitness value of $1$ mostly depends on the size of $\mathcal{D^A}$ and is relatively independent of the size of $\mathcal{D^C}$.
Intuitively this makes sense as we only need one entry $(\caseIdent,a) \in \mathcal{D^C}$ to deduce $a > b$, given that the newly received event is $(\caseIdent, b)$.
Even if case $\caseIdent$ is dropped at some point in time, and reinserted later, still information regarding the directly follows abstraction can be deduced.
However, if not enough space is reserved for the $\mathcal{D^A}$ data structure, then the data structure is incapable of storing the complete directly follows abstraction.

\subsection{Concept Drift}
\begin{figure}[tb]
	\centering
	\subfloat[Parallel to choice.]{ 
		\includegraphics[width=0.3\textwidth]{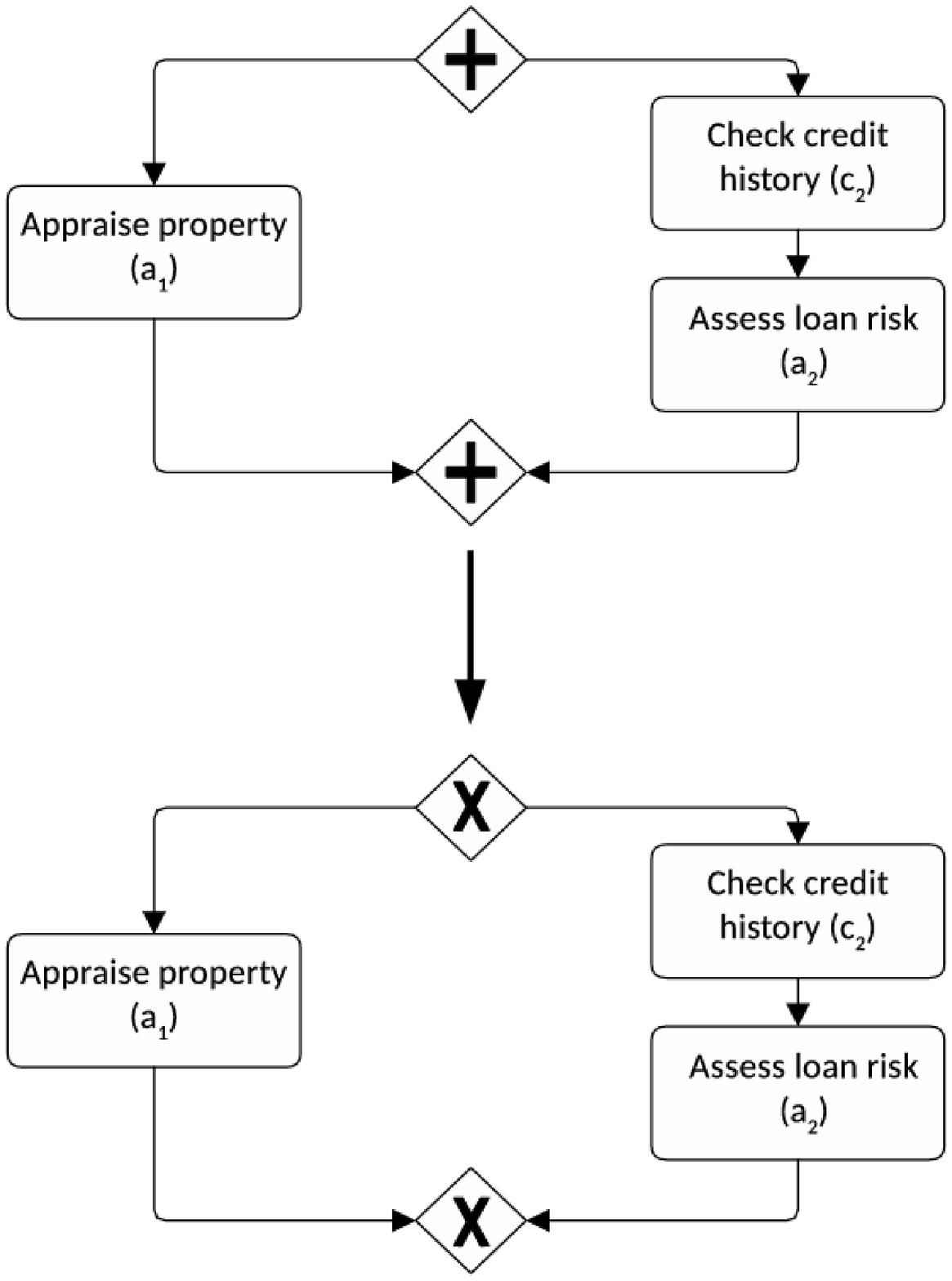}
		\label{fig:concept_drift_par_choi}
	}
	\subfloat[Choice to Parallel.]{
		\includegraphics[width=0.3\textwidth]{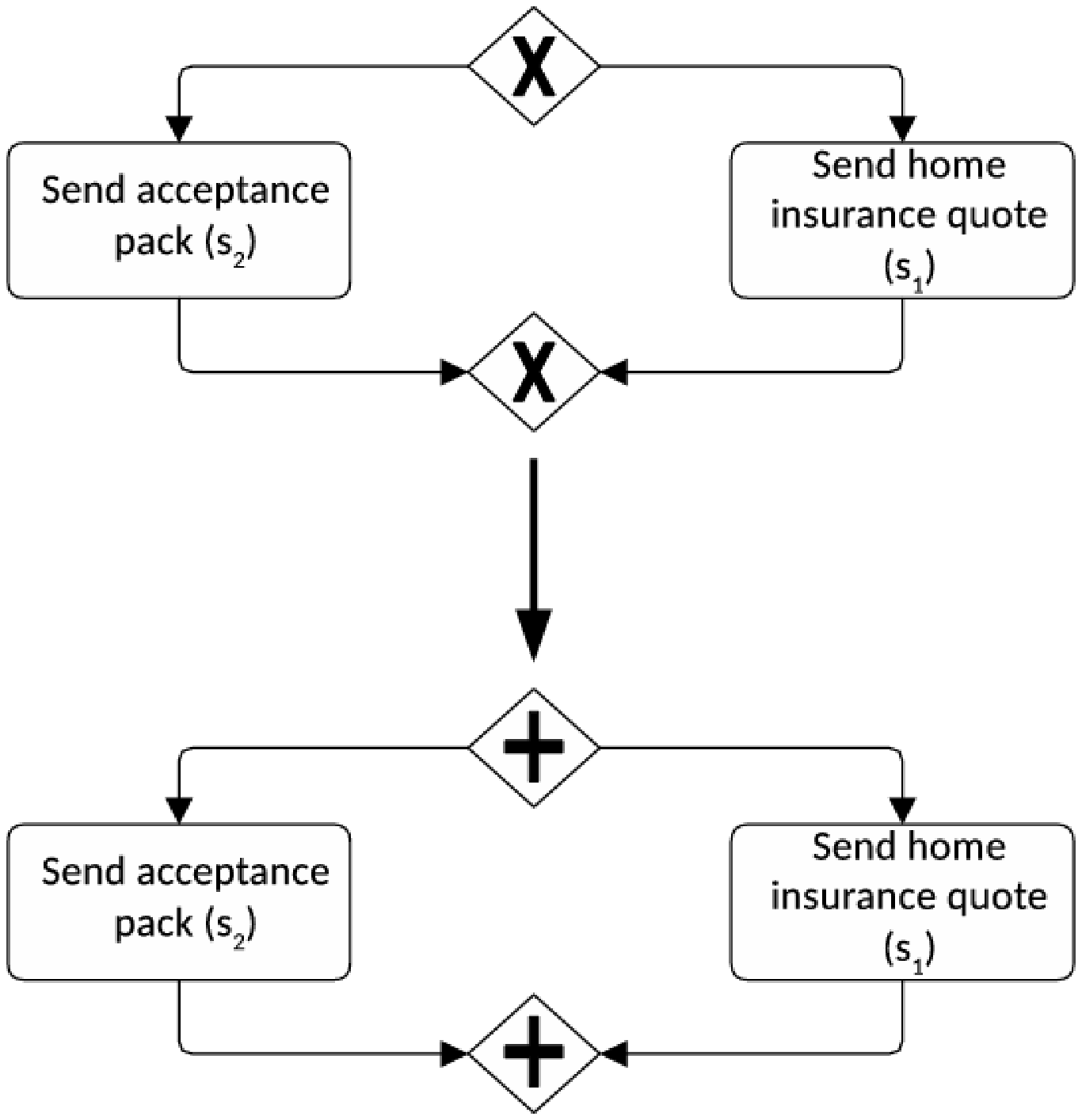}
		\label{fig:concept_drift_choi_par}
	}
	\caption{Changes made to the business process model presented in \autoref{fig:running_example_bpmn}.}
	\label{fig:concept_drift}
\end{figure}

In the previous experiments we focused on a process model that describes observed \textit{steady state} behaviour, i.e. the process model from which events are sampled does not change during the experiments.
In this section we assess to what extend the Inductive Miner based instantiation of the framework is able to handle \textit{concept drift}~\cite{DBLP:conf/aaai/SchlimmerG86, DBLP:journals/tnn/BoseAZP14}.
We focus on \textit{gradual drift}, i.e. the behaviour of the process model changes at some point in time, though the change is only applicable for new cases, already active cases follow the old behaviour.
In order to obtain a gradual drift, we manipulated the CPN simulation model of the process model presented in \autoref{fig:running_example_bpmn}.
The first five hundred cases that are simulated follow the original model.
All later cases are routed to a model in which we swap the parallel and choice structures within the model (\autoref{fig:concept_drift}).

\begin{figure}[tb]
	\centering
	\subfloat[$|\mathcal{D^C}| = 100$, $|\mathcal{D^A}| = 50$]{
		\includegraphics[width=0.5\textwidth]{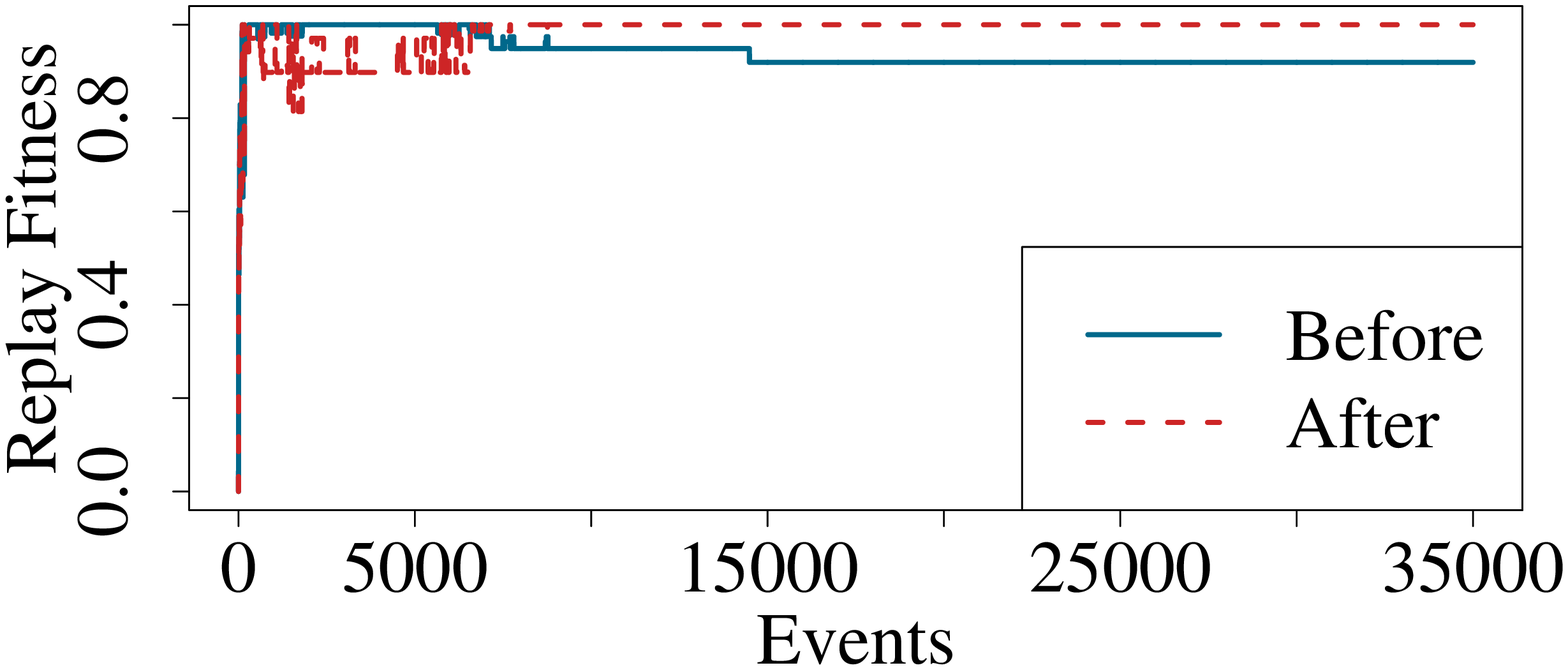}
		\label{fig:exp_4_concept_drift_50}
	}
	\subfloat[$|\mathcal{D^C}| = 100$, $|\mathcal{D^A}| = 100$]{
		\includegraphics[width=0.5\textwidth]{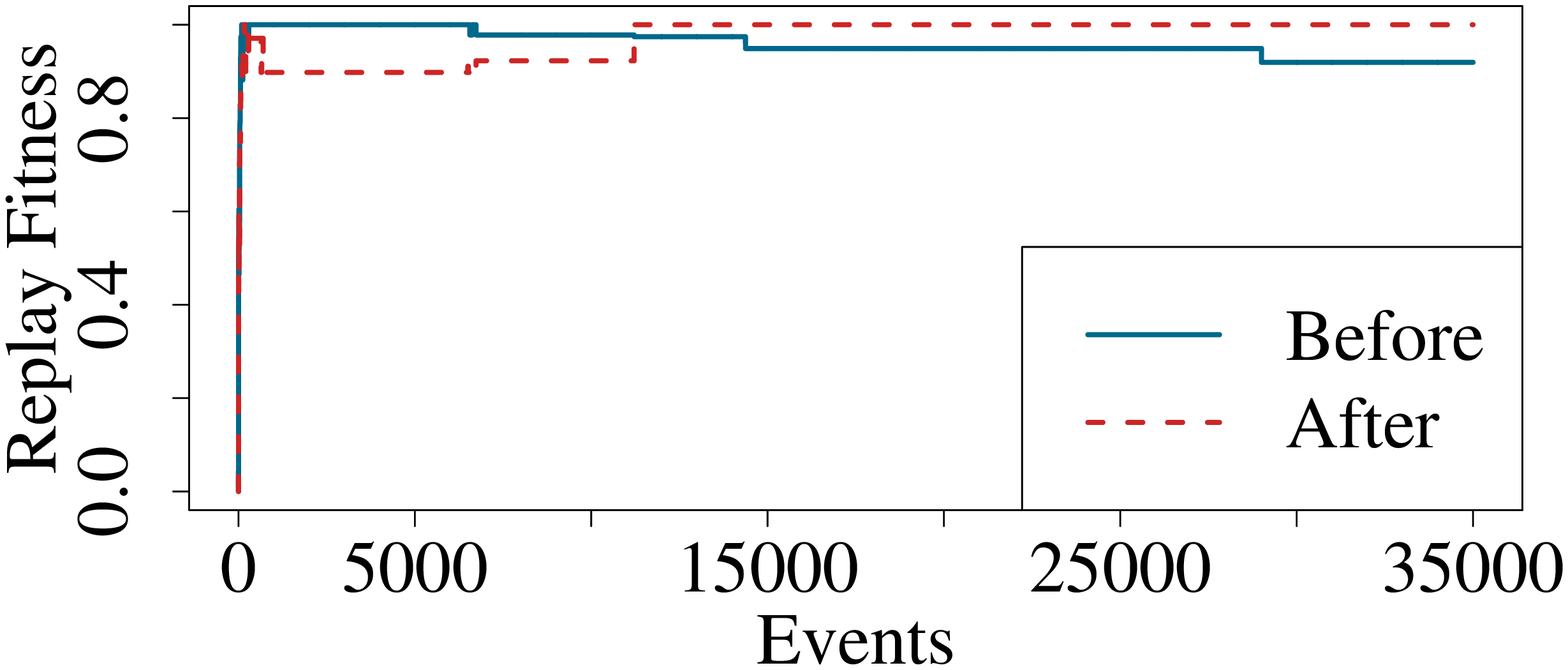}
		\label{fig:exp_4_concept_drift_100}
	}
	\caption{Replay Fitness measures for the Stream Inductive Miner, given an event stream containing concept drift.}
	\label{fig:exp_4_concept_drift}
\end{figure}

\autoref{fig:exp_4_concept_drift} depicts the results of applying the Inductive Miner on the described gradual drift.
In \autoref{fig:exp_4_concept_drift_50} we depict the results using data structure sizes $|\mathcal{D^C}| = 100$ and $|\mathcal{D^A}| = 50$ (Lossy Counting).
The blue solid line depicts the replay fitness w.r.t. an event log containing behaviour \textit{prior} to the drift, the red dashed line represents replay fitness w.r.t. an event log containing behaviour \textit{after} the drift.
We observe that the algorithm again needs some time to stabilize in terms of behaviour w.r.t. the pre-drift model.
Interestingly, at the moment that the algorithm seems to be stabilized w.r.t. the pre-drift model, the replay fitness w.r.t. the post-drift model fluctuates.
This indicates that the algorithm is not able to fully rediscover the pre-drift model, yet it produces a generalizing model which includes more behaviour, i.e. even behaviour that is part of the post-drift model.
The first event in the stream related to the new execution of the process, is the $6.415^{th}$ event.
Indeed, the blue solid line drops around this point in \autoref{fig:exp_4_concept_drift_50}.
Likewise, the red dashed line rapidly increase to value 1.0.
Finally, around event $15.000$ the replay fitness w.r.t. the pre-drift model stabilizes completely, indicating that the prior knowledge related to the pre-drift model is completely erased from the underlying data structure.
In \autoref{fig:exp_4_concept_drift_100} we depict results for the Inductive miner using sizes $|\mathcal{D^C}| = 100$ and $|\mathcal{D^A}| = 100$.
In this case we observe more stable behaviour, i.e. both the pre- and post-model behaviour stabilizes quickly.
Interestingly, due to the use of a bigger $k$-value of the Lossy Counting Algorithm, the drift is reflected longer in the replay fitness values.
Only after roughly the $30.000^{th}$ event the replay fitness w.r.t. the pre-drift model stabilizes.

\subsection{Performance Analysis}
\label{subsec:performance}
The main goal of the performance evaluation is to assess whether memory usage and processing times of the implementations are acceptable.
As the implementations are of a prototypical fashion, we focus on \textit{trends} in processing time and memory usage, rather than absolute performance measures.
For both processing time and memory usage, we expect stabilizing behaviour, i.e. over time we expect to observe some non-increasing asymptote.
If the processing time/memory usage keeps increasing over time this implies that we are potentially unable to handle data on the stream or need infinite memory.

\begin{figure}[tb]
	\centering
	\subfloat[Processing times in nanoseconds.]{
		\includegraphics[width=0.5\textwidth]{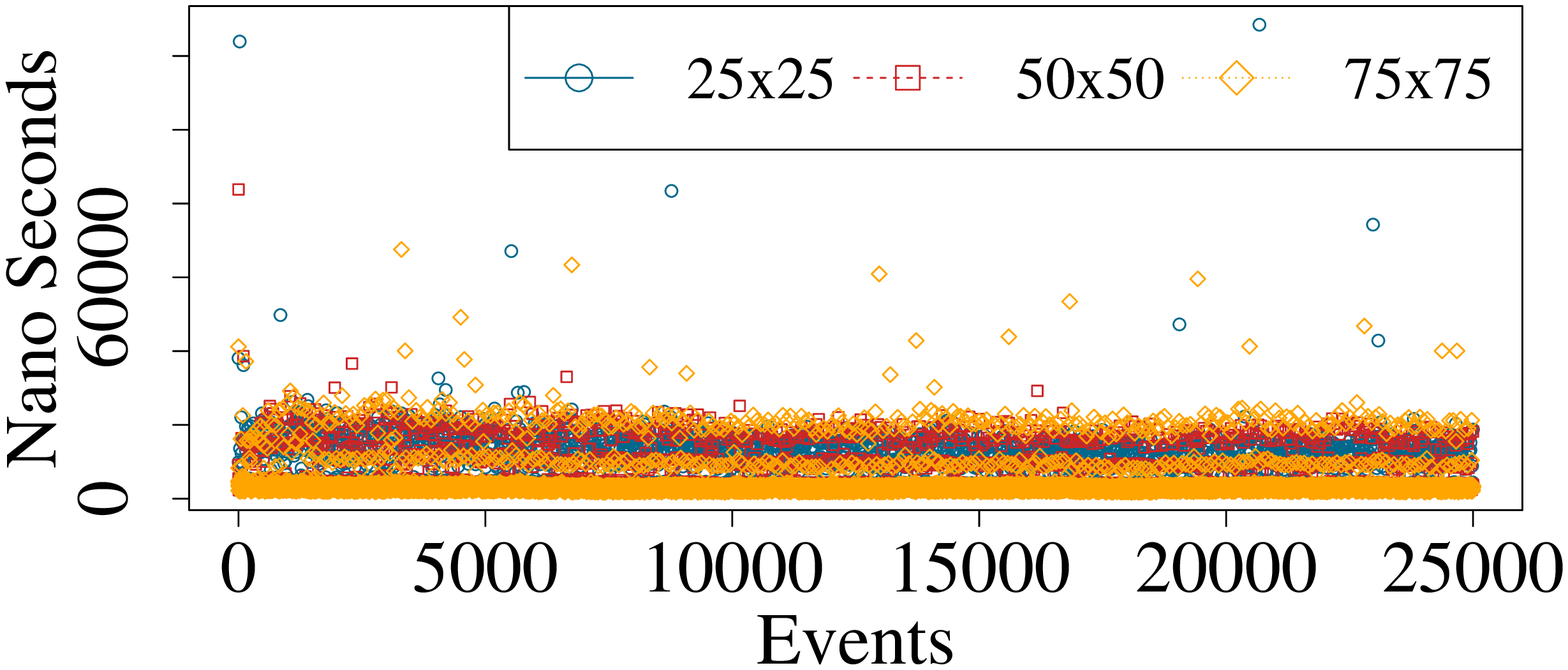}
		\label{fig:rt_25_50_75}
	}
	\subfloat[Memory usage in bytes.]{
		\includegraphics[width=0.5\textwidth]{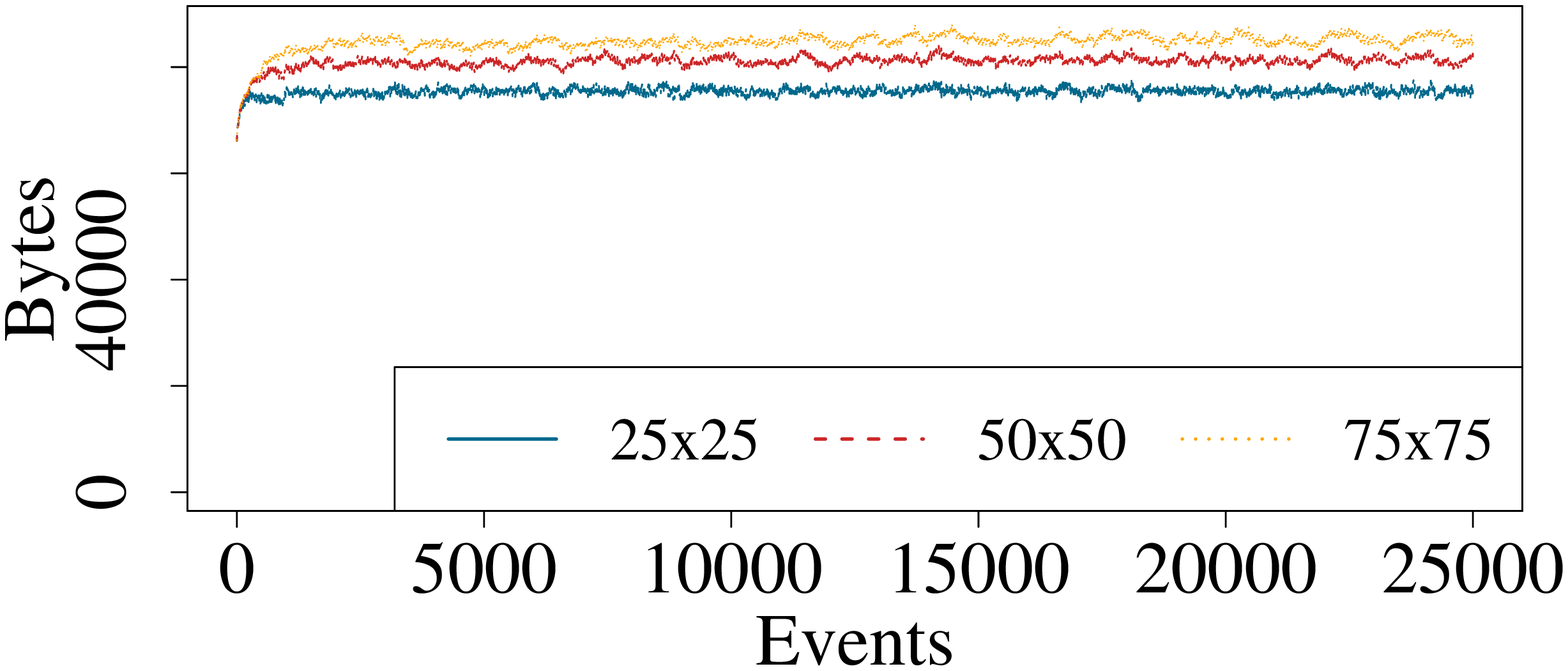}
		\label{fig:mem_25_50_75}
	}   
	\caption{Performance measurements based on the Stream Inductive Miner.}
	\label{fig:rt_mem}
\end{figure}

Within the experiment we measured the processing time and memory usage for handling the first 25.000 events emitted onto the stream.
We again use the Inductive Miner with Lossy Counting and varying window sizes (parameter $k$): $|\mathcal{D^C}| = 25$ and $|\mathcal{D^A}| = 25$, $|\mathcal{D^C}| = 50$ and $|\mathcal{D^A}| = 50$ and $|\mathcal{D^C}| = 75$, $|\mathcal{D^A}| = 75$ (represented in the Figures as 25x25, 50x50 and 75x75 respectively).
We measured the time the algorithm needs to update both $\mathcal{D^C}$ and $\mathcal{D^A}$.
The memory measured is the combined size of $\mathcal{D^C}$ and $\mathcal{D^A}$ in bytes.
The results of the experiments are depicted in \autoref{fig:rt_mem}.
Both figures depict the total number of events received on the x-axis.
In \autoref{fig:rt_25_50_75}, the processing time in nanoseconds is shown on the y-axis, whereas in \autoref{fig:mem_25_50_75}, the memory usage in bytes is depicted.
The aggregates of the experiments are depicted in Table~\ref{tab:rt_mem}.

\begin{table}[tb]
	\caption{Aggregate performance measures for the Stream Inductive Miner.}
	\label{tab:rt_mem}
	\centering
	\begin{tabular}{l c  c  c }
		\ & 25x25 & 50x50 & 75x75 \\
		\textit{Avg. processing time (ns.):} & 4.7167,77 & 3.866,45 & 3.519,22\\
		\textit{Stdev. processing time (ns.):} & 3.245,80 & 2.588,76 & 2.690,54\\
		& & & \\
		\textit{Avg. memory usage (byte):} & 75.391,75 & 81.013,60 & 84.695,86\\
		\textit{Stdev. memory usage (byte):} & 762,55 & 1.229,60 & 1724,98\\
	\end{tabular}
	\
\end{table}

As \autoref{fig:rt_25_50_75} shows, there is no observable increase in processing times as more events have been processed.
The average processing time seems to slightly decrease when the window size of the Lossy Counting data structure increases (see Table~\ref{tab:rt_mem}).
Intuitively this makes sense as a bigger window size of the Lossy Counting algorithm implies less frequent cleanup operations.

Like processing time, memory usage of the Lossy Counting data structures does not show an increasing trend (\autoref{fig:mem_25_50_75}).
In this case however, memory usage seems to increase when the window size of the Lossy Counting algorithm is bigger.
Again this makes sense, as less cleanup operations implies more active members within the data structures, and hence, a higher memory usage.

\section{Related Work}
\label{sec:related_work}
For a detailed overview of process mining we refer to~\cite{DBLP:books/sp/Aalst16}.
For an overview of models, techniques and algorithms in stream based mining and analysis, e.g. frequency approximation algorithms, we refer to~\cite{muthu_2005_data_streams,aggarwal_2007_data_streams,gama_2010_kd_from_streams}.
Little work has been done on the topic of stream-based process discovery, and, stream-based process mining in general.
The notion of \textit{streams of events} is not new, i.e. several fields study aspects related to streams of (discrete) events.
Compared to the field of Complex Event Processing (CEP)~\cite{DBLP:books/daglib/0024062}, the S-BAR architecture can be seen as an \textit{event consumer}, i.e. a decoupled entity that processes the events produced by the underlying system.
However, whereas the premise of CEP is towards the \textit{design} of event based systems and architectures, this work focuses on the \textit{behavioural analysis} of such systems.
The area of event mining~\cite{li_2015_event_mining}, focuses on gaining knowledge from  historical event/log data.
Although the input data is similar, i.e. streams of system events, the assumptions on the data source are different.
Within event mining, data mining techniques such as \textit{pattern mining}~\cite[Chpt. 4]{li_2015_event_mining} are used as opposed to techniques used within this paper, i.e. techniques discovering end-to-end process models with associated execution semantics.
Also, event mining includes methods for system monitoring, whereas the S-BAR architecture can serve as an enabler for business process monitoring and prediction.

To the best of the author's knowledge this paper is the first work that presents a generic architecture for the purpose of event stream based process discovery.
As such the work may be regarded as a generalization and standardization effort of some of the related work mentioned within this section.

In~\cite{burattin_2014_stream} an event stream based variant of the Heuristics Miner is presented.
The algorithm uses three internal data structures using both Lossy Counting~\cite{manku_2002_lossy} and Lossy Counting with Budget~\cite{martino_2013_lossy_budget}.
The authors use these structures to approximate a causal graph based on an event stream.
The authors additionally present a sliding window based approach.
%The authors propose an uncoupled scheme that only incorporates the $a > b$ abstraction.
%Both the $a >> b$ and $a >>> b$  abstractions are not considered.
%The work in this paper can be regarded as an abstraction, generalization and extension of the work in~\cite{burattin_2014_stream}.
%Similarly,~\cite{burattin_2014_stream} can be regarded as an instantiation of the architecture presented in this paper.
Recently an alternative data structure has been proposed based on prefix-trees~\cite{hassani_2015_stream_heuristics}.
In this work the authors deduce the directly follows abstraction directly from a prefix tree which is maintained in memory.
The main advantage of using the prefix-trees is the reduced processing time and usage of memory.
%The $\mathcal{D_C}$ data structure is represented by a set of pointers to nodes within the prefix tree.
%Although this method is designed to specifically finding a directly follows abstraction, it is straight forward to extend it to support length-$k$ distance relations.
In~\cite{redlich_2014_stream_ccm}, Redlich et al. design an event stream based variant of the CCM algorithm~\cite{redlich_2014_ccm}.
The authors identify the need to compute dynamic footprint information based on the event stream, which can be seen as the abstract representation used by CCM.
The dynamic footprint is translated to a process model using a translation step called \textit{Footprint Interpretation}.
%The Footprint Interpretation mainly reduces the overall complexity of the conventional CCM divide and conquer steps to allow the algorithm to be adapted in a streaming setting.
The authors additionally apply an ageing factor to the collected trace information to fade out the behaviour extracted from older traces.
Although the authors define event streams similarly to this paper the evaluation relies heavily on the concept of \textit{completed traces}.
%Similarly, [Boushaba et al.] recently developed an extension of their block structured miner ...
%Within this work the authors also assume the stream to be a stream of completed traces, rather than a stream of events.
In~\cite{burattin_2015_stream_declare} Burattin et al. propose an event stream based process discovery algorithm to discover declarative process models.
The structure described to maintain events and their relation to cases is comparable with the one used in~\cite{burattin_2014_stream}.
The authors present several declarative constraints that can be updated on the basis of newly arriving events instead of an event log consisting of full traces.

\section{Discussion}
\label{sec:discussion}
In this section we discuss interesting phenomena observed during experimentation which should be taken into account when adopting the architecture presented in this paper, and, in event stream based process discovery in general.
We discuss limitations w.r.t. the complexity of abstract representation computation and discuss the impact of the absence of trace initialization and termination information.

\subsection{Complexity of Abstract Representation Computation}
\label{subsec:restrictions}
There are limitations w.r.t. the algorithms we are able to adopt using abstract representations as basis.
This is mainly related to the computation of the abstract representation within the conventional algorithm.

As an example, consider the $\alpha^+$-algorithm~\cite{medeiros_2005_alpha_plus} which extends the original $\alpha$-Miner such that it is able to handle self-loops and length-1-loops.
For handling self-loops, the $\alpha^+$-algorithm traverses the event log and identifies activities that are within a self-loop.
Subsequently it removes these from the log and after that calculates the directly follows abstraction.
For example, if $L=[\langle a,b,c\rangle, \langle a,b,b,c \rangle]$, the algorithm will construct $L'= [\langle a,c \rangle]$ and compute directly follows metrics based on $L'$.

In a streaming setting we are able to handle this as follows.
Whenever we observe some activity $a$ to be in a self-loop and want to generate the directly follows abstraction, then for every $(a',a) \in \mathcal{D^A}$ and $(a,a'') \in \mathcal{D^A}$, s.t. $a \neq a'$ and $a \neq a''$, we deduce that $(a',a'')$ is part of the directly follows abstraction whereas $(a,a)$, $(a',a)$ and $(a,a'')$ are not.
Although this procedure approximates the directly follows relation on the event stream, a simple example shows that the relation is not always equal.

\begin{wrapfigure}{l}{0.5\textwidth}
	\centering
	\subfloat[Event log]{
		\begin{tikzpicture}[node distance = 1cm]
		\node (a) {$a$};
		\node (c) [right of = a] {$c$};
		\node (e) [below of = c] {$e$};
		\node (d) [right of = e] {$d$};
		
		\path[->]
		(a) edge node {} (c)
		(a) edge node {} (e)
		(e) edge node {} (d);
		\end{tikzpicture}
		\label{subfig:alpha_plus_stream_problem_log}
	} ~
	\subfloat[Event stream]{
		\begin{tikzpicture}[node distance = 1cm]
		\node (a) {$a$};
		\node (c) [right of = a] {$c$};
		\node (e) [below of = c] {$e$};
		\node (d) [right of = e] {$d$};
		
		\path[->]
		(a) edge node {} (c)
		(a) edge node {} (e)
		(e) edge node {} (d);
		
		\path[->, dashed]
		(a) edge node {} (d)
		(e) edge node {} (c);
		\end{tikzpicture}
		\label{subfig:alpha_plus_stream_problem_stream}
	}
	\caption{Two Abstract Representations.}
	\label{fig:alpha_plus_stream_problem}
\end{wrapfigure}
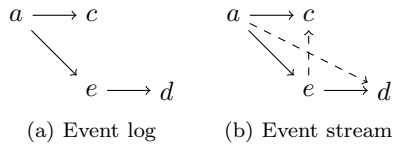
Imagine a process $\process = \{\langle a,b,b,c \rangle, \langle a,e,b,d \rangle\}$.
Clearly any noise-free event log over this process is just a multi-set over the two traces in $\process$.
In case of the conventional $\alpha^+$-algorithm, removing the $b$-activity leads to the two traces $\langle a,c \rangle$ and $\langle a,e,d \rangle$.
Consider the corresponding directly follows abstraction, depicted in \autoref{subfig:alpha_plus_stream_problem_log}.
Observe that all possible directly follows pairs that we are able to observe on any stream over $\process$ are: $(a,b), (a,e), (b,b), (b,c), (b,d), (e,b)$.
Applying the described procedure  yields the abstraction depicted in \autoref{subfig:alpha_plus_stream_problem_stream}.
Due to the information that is lost by only maintaining directly follows pairs, we deduce non-existing relations $(a,d)$ and $(e,c)$.

In general it is preferable to adopt an abstraction-based algorithm that constructs the abstract representation in \textit{one pass} over the event log.

\subsection{Initialization and Termination}
\label{sec:init_termination}
For the definitions presented in this paper, we abstract from trace initialization and/or termination, i.e. we do not assume the existence of explicit start/end events.
Apart from the technical challenges related to finding these events, i.e. as described in \autoref{sec:alpha_miner} regarding start/end activity sets used by the $\alpha$-Miner and Inductive Miner, this can have a severe impact on computing the abstract representation as well.

If we assume the existence and knowledge of unique start and end activities, adopting any algorithm to cope with this type of knowledge is trivial.
We only consider cases of which we identify a start event and we only remove knowledge related to cases of which we have seen the end event.
The only challenge is to cope with the need to remove an unfinished case due to memory issues, i.e. how to incorporate this deletion into the data structure/abstract representation that is approximated.

If we do not assume and/or know of the existence of start/end activities, whenever we encounter a case for which our data structure indicates that we have not seen it before, this case is identified as being a ``new case''.
Similarly, whenever we decide to drop a case from a data structure, we implicitly assume that this case has terminated.
Clearly, when there is a long period of inactivity, a case might be falsely assumed to be terminated.
If the case becomes active again, it is treated as a new case again.
The experiments reported on in \autoref{fig:exp_1_small_ds} show that in case of the directly follows abstraction, this type of behaviour has limited impact on the results.
However, in a more general sense, e.g. when approximating a prefix-closure on an event stream, this type of behaviour might be of greater influence w.r.t. resulting model.
The ILP Miner likely suffers from such errors and as a result produces models of inferior quality.

In fact, for the ILP Miner the concept of termination is of particular importance.
To guarantee a single final state of a process model, the ILP Miner needs to be aware of \textit{completed traces}.
This corresponds to explicit knowledge of when a case is terminated in an event stream setting.
Like in the case of initialization, the resulting models of the ILP miner are greatly influenced by a faulty assumption on case termination.

\section{Conclusion}
\label{sec:conclusion}
In this paper, we presented a generic architecture that allows for adopting existing process discovery algorithms in an event stream setting.
The architecture is based on the observation that many existing process discovery algorithms translate a given event log into some abstract representation and subsequently use this representation to discover a process model.
Thus, in an event stream based setting, it suffices to approximate the abstract representation using the event stream in order to apply existing process discovery algorithms to streams of events.
The exact behaviour present in the resulting process model greatly depends on the instantiation of the underlying techniques that approximate the abstract representation.

Several instantiations of the architecture have been implemented in the process mining tool-kits ProM and RapidProM.
We primarily focused on abstract representation approximations using algorithms designed for the purpose of frequent item mining on data streams.
We structurally evaluated and compared five different instantiations of the framework.
From a behavioural perspective we focused on the Inductive Miner as it grantees to produce sound workflow nets.
The experiments show that the instantiation is able to capture process behaviour originating from a steady state-based process.
Moreover, convergence of replay fitness to a stable value depends on parametrization of the internal data structure.
In case of concept drift, the size of the internal data structure of use impacts both model quality and the drift detection point.
We additionally studied the performance of the Inductive Miner instantiation.
The experiments show that both processing time of new events and memory usage are non-increasing as more data is received.

\paragraph{Future Work}
Within the experiments we chose to limit the use of internal data structure to the Lossy Counting based approach.
However, more instantiations, i.e. Frequent / Space Saving, are presented and implemented.
We plan to investigate the impact of several different designs of the internal data structures w.r.t. both behaviour and performance.

The architecture presented in this work focuses on approximating abstract representations and exploiting existing algorithms to discover a process model.
However, bulk of the work might be performed multiple times, i.e. several new events emitted to the stream might not change the abstract representation.
We therefore plan to conduct a study towards a completely incremental instantiation of the architecture, i.e. can we immediately identify whether new data changes the abstraction or even the resulting model?

Another interesting direction for future work is to go beyond control-flow discovery, i.e. can we lift conformance checking, performance analysis, etc. to the domain of event streams?
Moreover, in such cases we might need to store more information, i.e. store all attributes related to events within cases seen so far.
We plan to investigate the application of lossless/lossy compression of the data seen so far, i.e. using frequency distributions of activities/attributes to encode sequences in a compact manner.

%\begin{acknowledgements}
%\end{acknowledgements}

%\bibliography{bibliography_no_url}

\end{document}